\documentclass[a4paper,fleqn]{cas-sc}
\usepackage[square,numbers,sort&compress]{natbib}
\usepackage{blkarray}
\usepackage{tikz}
\usepackage[]{sidecap}

\newcommand{\xuparrow}[1]{%
  {\left\uparrow\vbox to #1{}\right.\kern-\nulldelimiterspace}
}

\begin{document}
\let\WriteBookmarks\relax
\def\floatpagepagefraction{1}
\def\textpagefraction{.001}
\shorttitle{Robust global tripartite entanglement in a  mixed spin-($1$,$1/2$,$1$) Heisenberg trimer   }
\shortauthors{H. Vargov\'a}

\title [mode = title]{Robust global tripartite entanglement in a  mixed spin-($1$,$1/2$,$1$) Heisenberg trimer  }

\author[1]{H. Vargov\'a}[
                        orcid=0000-0003-2000-7536]
\cormark[1]
\ead{hcencar@saske.sk}

\credit{Conceptualization, Methodology, Software validation, Formal analysis, Investigation, Resources, Data curation, Writing - Original draft preparation, Writing - review and editing, Visualization }

\affiliation[1]{organization={Institute of Experimental Physics, Slovak Academy of Sciences},
                addressline={Watsonova 47}, 
                city={Ko\v {s}ice},
                postcode={040 01}, 
                state={Slovakia}
                }

\cortext[cor1]{Corresponding author}

\begin{abstract}
We rigorously analyze the global tripartite entanglement in a Heisenberg trimer with mixed spins-($1$,$1/2$,$1$) under varying exchange couplings between dissimilar and identical  spins, magnetic fields, and temperatures. The global tripartite entanglement is quantified using the geometric mean of all three bipartite contributions, evaluated through negativity.  We precisely map the regions of parameter space in which the trimer system exhibits spontaneous global entanglement. In addition, we classify the nature of the tripartite entangled states based on the distribution of reduced bipartite negativities among the spin dimers in the trimer. 
We further examine the thermal stability of global tripartite entanglement throughout the full parameter space. Special attention is given to the theoretical prediction of thermal robustness in a real three-spin complex,  [Ni(bapa)(H$_2$0)]$_2$Cu(pba)(ClO$_4$)$_2$,  where bapa stands for bis($3$-aminopropyl)amine, and pba denotes $1$,$3$-propylenebis(oxamato), which serves as an experimental realization of the mixed-spin ($1$,$1/2$,$1$) Heisenberg trimer. Notably, global entanglement in this system is predicted to persist up to approximately $100$ K and magnetic fields approaching $210$ T.
Moreover, we uncover a thermally induced activation of robust global entanglement in regions where the ground state is biseparable. The magnitude of this thermal entanglement is remarkably high,  nearly reaching a value of $1/2$, which has not been reported before. Finally, we propose a connection between the theoretically predicted tripartite entanglement, quantified via negativity derived from the density matrix, and the quantities measured directly or indirectly from various experiments.

\end{abstract}

\begin{keywords}
mixed spin-($1$,$1/2$,$1$) Heisenberg trimer \sep global tripartite entanglement \sep negativity \sep Ni$_2$Cu molecular complexes
\end{keywords}

\maketitle

\section{Introduction}

Over the last four decades, nanomagnetism has become a highly active research area across various fields, including condensed matter physics, materials science, chemistry, computer science, etc~\cite{Kahn1,Sieklucka}. The interplay between long-range interactions, which are responsible for spin waves and skyrmionic behavior, and short-range interactions, which determine the discrete excitations of magnetic clusters, is one of key catalysts for the exotic properties of nanomagnets.  Another significant factor is the existence of strong magneto-electric coupling, which combined  with the low-dimensional nature of nanomagnets, makes them well-suited for use in spintronic devices~\cite{Kahn1, Gatteschi}. The ability to manipulate magnetic properties at the nanoscale is another notable feature, making molecular  magnets highly attractive  for applications in quantum computation and magnetic storage~\cite{Nielsen1,Tyagi}. Last but not least, nanomagnetic materials have garnered significant interest due to the non-local nature of quantum correlations, known as entanglement~\cite{EPR,Bohr,Vedral}.

Entanglement is a fundamental property of composite quantum systems that prevents us from describing the state of any single constituent independently of the rest of the system. In an entangled system, the state of each part is intrinsically linked to the state of the others, so knowing the state of one part allows us to immediately infer the state of its counterpart without the need for measurement. This interconnectedness, coupled with the absence of measurement, has become a cornerstone in advancing secure communication technologies through quantum cryptography~\cite{Bechmann}. Simultaneously, entanglement plays a pivotal role in the realm of quantum computing, where the absence of measurement is responsible for substantial increases in computational speed~\cite{Arnesen}.

The smallest magnetic cluster in which entanglement can be observed is the magnetic dimer, consisting of two spins. This type of  entanglement is known  as pairwise or bipartite entanglement and is well-understood at present. For more information, we recommend excellent review papers~\cite{Amico,Horodecki1}.  In addition to these, several other scientific studies have focused on identifying  appropriate intrinsic or extrinsic mechanisms that  enable the manipulation of entanglement strength. Special emphasis is placed on stabilizing thermal entanglement at higher temperatures (ideally around room temperature). Various mechanisms, such as (non)uniform external magnetic fields, single-ion anisotropy, exchange anisotropy, and spin diversity, have been discussed in detail. For more details, see  Refs.~\cite{Gong,Mahdavifar,Sun,Khedif,Zhou2015, Li,Solano,Guo2010,Abliz,Cenci2020,Cenci2021,Cenci2022}, and the references therein. 
Beyond bipartite entanglement, there exists global (multipartite) entanglement, which   occurs simultaneously between each pair of spins in a multispin magnetic cluster~\cite{Carvalho}. Despite extensive research efforts, a comprehensive understanding of global entanglement remains an open and intensively investigated task, primarily motivated by the need to better understand interactions between various registers of a quantum computer. Further motivation stems from the current recognition that higher spin or a greater number of spin constituents in small magnetic clusters can produce non-trivial macroscopic properties~\cite{Sieklucka}, and therefore, an increase in entanglement could also be anticipated.

It should be mentioned that there are a few interesting studies dealing with  entanglement in multipartite spin systems, well described by the Heisenberg model. Many of these studies focus on particles with identical spin magnitudes, e.g., Refs.~\cite{Su2022,Ou,Benabdallah,Tian,Szalowski,Zad2022,
Karlova2023,Hawary,Khan2018,Zad2025}, but some also consider small mixed-spin clusters, e.g., Refs.~\cite{Zad2016,Vargova2024a,Ahami,Adamyan2024,Adamyan2024a,Ghannadan2025,Vargova2025}.  The limited number of studies focusing on the mixed-spin alternatives is due to the higher mathematical complexity. Nevertheless, there is a plethora of real multinuclear complexes with various magnetic ions in a magnetic core, such as complexes with molecular units based on Ni$_2$Cu, Ni$_2$Co, Fe$_2$Ni, Fe$_2$Mn, Fe$_2$Ni$_2$,  Fe$_2$Cu$_2$, and others~\cite{Tercero,Ribas,Pei,Podlesnyak,Parkin,Rodriguez,Park,Wu,Ribas1,Nakamura,
Osa,Lou,Barrios,Barclay}. 
In the simplest multipartite systems (the tripartite one), global entanglement is usually quantified by the geometric mean of all bipartite entanglements~\cite{Benabdallah,Tian,Khan2018,Vargova2024a,
Ghannadan2025,Zad2025}, though some studies prefer the arithmetic mean instead~\cite{Su2022,Ou}. Additionally, the definition of tripartite concurrence is also used to quantify the degree of tripartite entanglement~\cite{Zad2016,Hawary}. It should be noted that besides the aforementioned techniques, other methods exist for quantifying the degree of tripartite entanglement, such as the Coffman, Kundu, and Wootters (CKW) inequality relation~\cite{CKW} and the concurrence triangle method~\cite{Xie2021}.  However, these procedures have so far been utilized to estimate the strength of tripartite entanglement in specific states like the Greenberger-Horne-Zeilinger (GHZ) and W states only.

In the present paper, we delve into the complex problem of multipartite entanglement by studying the smallest tripartite system with mixed spins. We focus on the configuration of spins-($1$,$1/2$,$1$), where the contribution from the spin-$1$ dimer could lead to a stronger enhancement of global tripartite entanglement compared to the simpler spin-($1/2$,$1$,$1/2$) alternative, due to the higher total spin of the selected dimer. Additionally, we expect that the higher effective spin of the selected spin-$1$ dimer could significantly stabilize global entanglement at higher temperatures. It should be noted that while such spin configurations are rare in real molecular complexes, there are a few trinuclear complexes with the spin configuration ($1$,$1/2$,$1$).  Examples include  oxamato-bridged heterotrinuclear Ni(II)Cu(II)Ni(II) complexes~\cite{Tercero,Ribas,Pei}, polycrystalline sample Ca$_3$CuNi$_2$(PO$_4$)$_4$~\cite{Podlesnyak}, and bimetallic nickel complexes of a bridging verdazyl radical~\cite{Barclay}. Although all of these complexes are nearly linear three-spin systems, we analyze a more general case with varying exchange interaction values between terminal ion spins, as a prerequisite for synthesizing novel and interesting complexes.

The paper is organized as follows. In Section~\ref{model}, we define an appropriate quantum model and introduce the relevant quantities to measure  entanglement in tripartite systems. Our definition is based on measuring the bipartite negativities of the entire (non-reduced) system, established from the explicit form of the respective density matrix. The obtained results are collected in Section~\ref{results}, where various types of tripartite entangled states, categorized according to the classification by Sab\'{i}n and Garc\'{i}a-Alcaine~\cite{Sabin}, are identified based on variations in model parameters. The thermal stability of tripartite negativity is also analyzed with respect to variations in the external magnetic field and exchange interactions. For a specific set of model parameters corresponding to a real spin complex, we predict the presence of global entanglement at relatively high temperatures and magnetic field strengths. Additionally, we discuss in detail the most important observation related to the thermal activation of global tripartite entanglement in biseparable ground states characterized by zero global tripartite entanglement at zero temperature. Despite expecting only a slight activation of global entanglement on the order of $\sim$0.01, we identify surprisingly robust global negativity achieving maximal magnitudes around $\sim$0.488. In the last subsection of Section~\ref{results}, we briefly discuss the relation between theoretical predictions of entanglement and their experimental evidence through direct and indirect measurements, including standard magnetometry experiments and inelastic neutron scattering experiments. Finally, some concluding remarks are presented in Section~\ref{conclusion}. Relevant technical details are provided in the Appendix.

\section{\label{model} Model and Method}
Let us consider the spin-($1$,$1/2$,$1$) Heisenberg trimer in the presence of an external magnetic field $B$, defined through  the Hamiltonian:
\allowdisplaybreaks
\begin{align}
\hat{\cal H}&=J\left(\hat{\bf S}_{1}
\!+\!\hat{\bf S}_{2}\right)\cdot\hat{\boldsymbol\mu}
\!+\! J_1\hat{\bf S}_{1}\cdot\hat{\bf S}_{2}
\!-\!h\left(\hat{S}^z_{1}\!+\!\hat{S}^z_{2}\!+\!\hat{\mu}^z\right).
\label{eq1}
\end{align} 
The spin operators $\hat{\bf S}_{1}$ and $\hat{\bf S}_{2}$ represent two identical Heisenberg spins with a magnitude $S_i=$1, while the spin operator $\hat{\boldsymbol\mu}$ represents a dissimilar Heisenberg spin with a magnitude of $\mu=$1/2.  $\hat{S}_i^z$  and $\hat{\mu}^z$ denote the $z$-component of the respective spin operators. $J$ and $J_1$ represent the exchange coupling constants between dissimilar and identical spins, respectively.
In general, both exchange coupling constants can be ferromagnetic or antiferromagnetic. The effect of an external magnetic field $B$,  applied along the $z$-direction, is described by the last term of Eq.~\eqref{eq1} through the constant $h=g\mu_BB$, where $g$ is the Land\'e g-factor and $\mu_B$ is the Bohrn magneton.

 An exhaustive study of the entanglement in a tripartite system is  conducted at two levels.  First, a comprehensive analysis of the reduced bipartite entanglement between two spins, $A$ and $B$, within the tripartite system $ABC$ is presented. Subsequently, the global tripartite entanglement is quantified, and the results of the reduced bipartite analysis are used to classify different types of tripartite entanglement.

The reduced bipartite entanglement between spins $A$ and $B$ is quantified using the bipartite negativity~\cite{Vidal}
\begin{align}
{\cal N}_{A|B}\!=\!{\cal N}(\hat{\rho}_{{AB}}^{T_A})\!=\!\sum_{(\lambda_{A|B})_i<0}\vert (\lambda_{A|B})_i\vert,
\label{eq2}
\end{align}
which is defined as the sum of the absolute values of all negative eigenvalues $|(\lambda_{A|B})_i|$ of the partially transposed density matrix $\hat{\rho}_{AB}^{T_{A}}$.
The   density matrix is obtained from the original one, $\hat{\rho}_{ABC}=\tfrac{1}{\cal Z}\sum_i\exp(-\beta\varepsilon_i)\vert\psi_i\rangle\langle\psi_i\vert$, by tracing out the degrees of freedom of the remaining spin C, yielding $\hat{\rho}_{AB}=\tfrac{1}{\cal Z}\sum_i {\rm Tr}_{C}\exp(-\beta\varepsilon_i)\vert\psi_i\rangle\langle\psi_i\vert$. Here, ${\cal Z}=\sum_i\exp(-\beta\varepsilon_i)$ is the partition function, $\varepsilon_i$ and $\vert \psi_i\rangle$ are an eigenvalues and eigenvectors of the model Hamiltonian~\eqref{eq1}, and $\beta$ is the inverse temperature. A partial transposition can be performed with respect to either subsystem $A$ or $B$, yielding the same value of bipartite negativity. The chosen subsystem is indicated by the upper index  $T_{A} (T_{B})$.  
Since one of the triplet spins is reduced, we refer to the resulting transposed density matrix, $\hat{\rho}_{AB}^{T_{A}}$, as the reduced transposed density matrix, and to ${\cal N}_{A|B}$ as the reduced bipartite negativity.
 Due to the inhomogeneity of spins in a mixed spin-($1$,$1/2$,$1$) Heisenberg trimer, there exist  two inequivalent reduced bipartite negativities, ${\cal N}_{\mu|S_1}\equiv{\cal N}_{\mu|S_2}$ and ${\cal N}_{S_1|S_2}$. According to the definition~\eqref{eq2},  both reduced bipartite negativities are non-negative functions. A zero magnitude indicates a separated (unentangled) state, while a positive value reflects the strength of bipartite entanglement.

The global tripartite entanglement among tree spins $A$, $B$, and $C$ is determined using the geometric mean of all three bipartite negativities~\cite{Sabin}
\begin{align}
{\cal N}_{ABC}\!=\!\sqrt[3]{{\cal N}_{A|BC} {\cal N}_{B|AC} {\cal N}_{C|AB} }.
\label{eq3}
\end{align}
The subscript $A|BC$ ($B|AC$, $C|AB$) indicates the type of bisection, where the spin on the left-hand side of the vertical line (e.g., $A$) represents one part, and the pair of spins on the right-hand side (e.g., $BC$) represents the other part of the tripartite system.  Each bipartite negativity, such as ${\cal N}_{A|BC}$, is computed directly using the standard definition~\cite{Vidal}
\begin{align}
{\cal N}_{A|BC}\!=\!{\cal N}(\hat{\rho}_{{ABC}}^{T_A})\!=\!\sum_{(\lambda_{A|BC})_i<0}\left\vert(\lambda_{A|BC})_i\right\vert,
\label{eq4}
\end{align}
where $(\lambda_{A|BC})_i$ are the eigenvalues of the partially transposed density matrix $\hat{\rho}_{{ABC}}^{T_A}$. Due to spin inhomogeneity, there exist two non-equivalent bipartite negativities, namely  ${\cal N}_{\mu|S_1S_2}$ and ${\cal N}_{S_1|\mu S_2}\!=\!{\cal N}_{S_2|\mu S_1}$, which contribute to the global tripartite negativity. Thus, the global tripartite negativity is given by  ${\cal N}_{\mu S_1S_2}\!=\!\sqrt[3]{{\cal N}_{\mu|S_1S_2} {\cal N}^2_{S_1|\mu S_2} }$.

Finally, we note that all computational details and the complete set of analytical results are collected in the Appendix.
\section{\label{results} Results and discussion}

\subsection{\label{Zero-temperature behavior} Zero-temperature behavior}

Diagonalizing the Hamiltonian~\eqref{eq1} in the standard basis  $\{\vert \mu^z,S_1^z,S_2^z\rangle\}$, where $\mu^z=\pm 1/2$ and $S_i^z=\{0,\pm 1\}$, yields a  complete set of  eigenvalues and corresponding eigenvectors, which are summarized in  Tab.~\ref{tabA1}.
\begin{table}[h!]
\caption{The complete list of eigenvalues and corresponding eigenvectors of the model~\eqref{eq1}, obtained in the standard basis $\vert \mu^z,S_1^z,S_2^z\rangle$, where $\mu^z=\pm1/2$ and $S_i^z=\{0,\pm1\}$.}
\label{tabA1}
\resizebox{1\textwidth}{!}{
\begin{tabular}{|c| l|l | }
\hline
$S_t$ & Eigenvalues $\varepsilon_{S_{t},S_{t}^z}$ & Eigenvectors $\vert S_t,S_t^z\rangle$\\
\hline
$\tfrac{5}{2}$&$\varepsilon_{\tfrac{5}{2},\pm\tfrac{5}{2}}=J+J_1\mp\tfrac{5}{2}h$ & $\vert \tfrac{5}{2},\pm\tfrac{5}{2}\rangle=\vert \pm\tfrac{1}{2},\pm1,\pm1\rangle$\\
&$\varepsilon_{\tfrac{5}{2},\pm\tfrac{3}{2}}=J+J_1\mp\tfrac{3}{2}h$ & $\vert \tfrac{5}{2},\pm\tfrac{3}{2}\rangle=\frac{\sqrt{2}}{\sqrt{5}}\left(\vert \pm\tfrac{1}{2},\pm1,0\rangle+\vert \pm\tfrac{1}{2},0,\pm1\rangle+\tfrac{1}{\sqrt{2}}\vert \mp\tfrac{1}{2},\pm1,\pm1\rangle\right)$\\
&$\varepsilon_{\tfrac{5}{2},\pm\tfrac{1}{2}}=J+J_1\mp\tfrac{h}{2}$ & $ \vert \tfrac{5}{2},\pm\tfrac{1}{2}\rangle=\frac{1}{\sqrt{10}}\left(\vert \pm\tfrac{1}{2},\pm1,\mp1\rangle+2\vert \pm\tfrac{1}{2},0,0\rangle+\vert \pm\tfrac{1}{2},\mp1,\pm1\rangle\right)+\frac{1}{\sqrt{5}}\left(\vert \mp\tfrac{1}{2},\pm1,0\rangle+\vert \mp\tfrac{1}{2},0,\pm 1\rangle\right)$\\
\hline
$\tfrac{3}{2}$&$\varepsilon_{\tfrac{3}{2},\pm\tfrac{3}{2}}^{\rm I}=\tfrac{J}{2}-J_1\mp\tfrac{3}{2}h$ & $\vert \tfrac{3}{2},\pm\tfrac{3}{2}\rangle^{\rm I}=\mp\frac{1}{\sqrt{2}}\left(\vert \pm\tfrac{1}{2},\pm1,0\rangle-\vert \pm\tfrac{1}{2},0,\pm1\rangle\right)$\\
&$\varepsilon_{\tfrac{3}{2},\pm\tfrac{3}{2}}^{\rm II}=-\tfrac{3}{2}J+J_1\mp\tfrac{3}{2}h$ & $\vert \tfrac{3}{2},\pm\tfrac{3}{2}\rangle^{\rm II}=\mp\frac{1}{\sqrt{10}}\left(\vert \pm\tfrac{1}{2},\pm1,0\rangle+\vert \pm\tfrac{1}{2},0,\pm1\rangle-2\sqrt{2}\vert \mp\tfrac{1}{2},\pm1,\pm1\rangle\right)$\\
&$\varepsilon_{\tfrac{3}{2},\pm\tfrac{1}{2}}^{\rm I}=\tfrac{J}{2}-J_1\mp\tfrac{h}{2}$ & $ \vert \tfrac{3}{2},\pm\tfrac{1}{2}\rangle^{\rm I}=\mp\frac{1}{\sqrt{3}}\left(\vert \pm\tfrac{1}{2},\pm1,\mp1\rangle-\vert \pm\tfrac{1}{2},\mp1,\pm1\rangle\right)\mp\frac{1}{\sqrt{6}}\left(\vert \mp\tfrac{1}{2},\pm1,0\rangle-\vert \mp\tfrac{1}{2},0,\pm 1\rangle\right)$\\
&$\varepsilon_{\tfrac{3}{2},\pm\tfrac{1}{2}}^{\rm II}=-\tfrac{3}{2}J+J_1\mp\tfrac{h}{2}$ & $\vert \tfrac{3}{2},\pm\tfrac{1}{2}\rangle^{\rm II}=\frac{1}{\sqrt{15}}\left(\vert \pm\tfrac{1}{2},\pm1,\mp1\rangle+2\vert \pm\tfrac{1}{2},0,0\rangle+\vert \pm\tfrac{1}{2},\mp1,\pm1\rangle\right)\mp\tfrac{\sqrt{3}}{\sqrt{10}}\left(\vert \mp\tfrac{1}{2},\pm1,0\rangle+\vert \mp\tfrac{1}{2},0,\pm 1\rangle\right)$\\
\hline
$\tfrac{1}{2}$&$\varepsilon_{\tfrac{1}{2},\pm\tfrac{1}{2}}^{\rm I}=-2J_1\mp\tfrac{h}{2}$ & $\vert \tfrac{1}{2},\pm\tfrac{1}{2}\rangle^{\rm I}=\frac{1}{\sqrt{3}}\left(\vert \pm\tfrac{1}{2},\pm1,\mp1\rangle-\vert \pm\tfrac{1}{2},0,0\rangle+\vert \pm\tfrac{1}{2},\mp1,\pm1\rangle\right)$\\
&$\varepsilon_{\tfrac{1}{2},\pm\tfrac{1}{2}}^{\rm II}=-J-J_1\mp\tfrac{h}{2}$ & $\vert \tfrac{1}{2},\pm\tfrac{1}{2}\rangle^{\rm II}=\frac{1}{\sqrt{6}}\left(\vert \pm\tfrac{1}{2},\pm1,\mp1\rangle-\vert \pm\tfrac{1}{2},\mp1,\pm1\rangle\right)-\frac{1}{\sqrt{3}}\left(\vert \mp\tfrac{1}{2},\pm1,0\rangle-\vert \mp\tfrac{1}{2},0,\pm 1\rangle\right)$\\
\hline
\end{tabular}
}
\end{table}
The obtained eigenvalues, $\varepsilon_{S_t,S_t^z}$, and  eigenvectors, $\vert S_t,S_t^z\rangle$, are classified according to the eigenvalues of two composite spin operators: the total spin operator  $\hat{S}_t=\hat{\mu}+\hat{S}_1+\hat{S}_2$ and its $z$-component $\hat{S}^z_t$.
   
By minimizing the energy across the full parameter space, we identify four pure, non-degenerate ferrimagnetic ground states:  $\vert \tfrac{1}{2},\tfrac{1}{2}\rangle^{\rm I}$, $\vert \tfrac{1}{2},\tfrac{1}{2}\rangle^{\rm II}$, $\vert  \tfrac{3}{2},\tfrac{3}{2}\rangle^{\rm I}$, and $\vert  \tfrac{3}{2},\tfrac{3}{2}\rangle^{\rm II}$, along with one pure non-degenerate ferromagnetic ground state,  $\vert \tfrac{5}{2},\tfrac{5}{2}\rangle$. The structure of the corresponding ground-state phase diagram is depicted in Fig.~\ref{fig1}. The two ferrimagnetic ground states, $\vert \tfrac{1}{2},\tfrac{1}{2}\rangle^{\rm II}$, $\vert  \tfrac{3}{2},\tfrac{3}{2}\rangle^{\rm II}$, are stable exclusively for antiferromagnetic exchange coupling $J > 0$, whereas the exchange coupling $J_1$ can be either antiferromagnetic or ferromagnetic.  The remaining three ground states appear regardless of whether $J$ is ferromagnetic or antiferromagnetic. Consequently, the phase diagram for $J < 0$ can be regarded as a simplified subset of the $J > 0$ case. Therefore, all conclusions derived for $J > 0$ and $J_1/|J| > 1$ can be generalized to the ferromagnetic regime $J < 0$, with appropriate attention to the modified phase boundaries.

It is worth mentioning that the analyzed model~\eqref{eq1} can also exhibit additional degenerate ground states (mixed states) within very narrow regions of the parameter space. Along the boundary lines
\begin{align}
\begin{array}{lll}
\vert \tfrac{5}{2},\tfrac{5}{2}\rangle\;-\vert  \tfrac{3}{2},\tfrac{3}{2}\rangle^{\rm I}\;: \; h/|J|=2J_1/|J|+1/2,&&\vert \tfrac{5}{2},\tfrac{5}{2}\rangle\;\;-\vert  \tfrac{3}{2},\tfrac{3}{2}\rangle^{\rm II}: \; h/|J|=5/2,\\
\vert \tfrac{3}{2},\tfrac{3}{2}\rangle^{\rm I}-\vert  \tfrac{1}{2},\tfrac{1}{2}\rangle^{\rm I}\;: \; h/|J|=J_1/|J|+1/2,&&\vert \tfrac{3}{2},\tfrac{3}{2}\rangle^{\rm II}-\vert  \tfrac{1}{2},\tfrac{1}{2}\rangle^{\rm II}: \; h/|J|=2J_1/|J|-1/2,\\
\vert \tfrac{3}{2},\tfrac{3}{2}\rangle^{\rm I}-\vert  \tfrac{3}{2},\tfrac{3}{2}\rangle^{\rm II}: \; h/|J|=1,&&\vert \tfrac{1}{2},\tfrac{1}{2}\rangle^{\rm I}\;-\vert  \tfrac{1}{2},\tfrac{1}{2}\rangle^{\rm II}: \; h/|J|=1,
\end{array}
\end{align}
we identify two-fold degenerate ferrimagnetic ground states, where two eigenstates coexist with identical energies
\begin{align}
\vert {\rm FI}\rangle^{\rm I}_{\tfrac{5}{2}-\tfrac{3}{2}}&=\left\{
\begin{array}{ll}
\vert \tfrac{5}{2},\tfrac{5}{2}\rangle\\
\vert \tfrac{3}{2},\tfrac{3}{2}\rangle^{\rm I}
\end{array}
\right.,
\;\;\;\;\vert {\rm FI}\rangle^{\rm II}_{\tfrac{5}{2}-\tfrac{3}{2}}=
\left\{\begin{array}{ll}
\vert \tfrac{5}{2},\tfrac{5}{2}\rangle\\
\vert \tfrac{3}{2},\tfrac{3}{2}\rangle^{\rm II}
\end{array}
\right.,
\;\;\;\;\;\;
\vert {\rm FI}\rangle_{\tfrac{3}{2}-\tfrac{3}{2}}=\left\{
\begin{array}{ll}
\vert \tfrac{3}{2},\tfrac{3}{2}\rangle^{\rm I}\\
\vert \tfrac{3}{2},\tfrac{3}{2}\rangle^{\rm II}
\end{array}
\right.,
\nonumber\\
\vert {\rm FI}\rangle^{\rm I}_{\tfrac{3}{2}-\tfrac{1}{2}}&=\left\{
\begin{array}{ll}
\vert \tfrac{3}{2},\tfrac{3}{2}\rangle^{\rm I}\\
\vert \tfrac{1}{2},\tfrac{1}{2}\rangle^{\rm I}
\end{array}
\right.,
\;\;\;\;\vert {\rm FI}\rangle^{\rm II}_{\tfrac{3}{2}-\tfrac{1}{2}}=\left\{
\begin{array}{ll}
\vert \tfrac{3}{2},\tfrac{3}{2}\rangle^{\rm II}\\
\vert \tfrac{1}{2},\tfrac{1}{2}\rangle^{\rm II}
\end{array}
\right.,
\;\;\;\;\;\;\vert {\rm FI}\rangle_{\tfrac{1}{2}-\tfrac{1}{2}}=\left\{
\begin{array}{ll}
\vert \tfrac{1}{2},\tfrac{1}{2}\rangle^{\rm I}\\
\vert \tfrac{1}{2},\tfrac{1}{2}\rangle^{\rm II}
\end{array}
\right..
\end{align}
Additionally, four-fold and three-fold degenerate ferrimagnetic ground states may exist at the intersection points of the boundary lines and the isotropic limit $J_1/|J|=1$
\begin{align}
\vert {\rm FI}_1\rangle^{1}&=\left\{
\begin{array}{l}
\vert \tfrac{1}{2},\tfrac{1}{2}\rangle^{\rm I}\\
\vert \tfrac{1}{2},\tfrac{1}{2}\rangle^{\rm II}\\
\vert \tfrac{3}{2},\tfrac{3}{2}\rangle^{\rm I}\\
\vert \tfrac{3}{2},\tfrac{3}{2}\rangle^{\rm II}
\end{array}
\right.,
\;\;\;
\vert {\rm FI}_1\rangle^{2}=\left\{
\begin{array}{l}
\vert \tfrac{3}{2},\tfrac{3}{2}\rangle^{\rm I}\\
\vert \tfrac{3}{2},\tfrac{3}{2}\rangle^{\rm II}\\
\vert \tfrac{5}{2},\tfrac{5}{2}\rangle\end{array}
\right..
\end{align}
Finally, in the absence of a magnetic field ($h/|J|=0$), nine multi-fold degenerate ground states are favored.\\
At $J>0$:
\begin{align}
\vert {\rm FI}^+_0\rangle^1&=\left\{
\begin{array}{l}
\vert \tfrac{3}{2},\tfrac{1}{2}\rangle^{\rm II}\\
\vert \tfrac{3}{2},\tfrac{3}{2}\rangle^{\rm II}\\
\vert \tfrac{3}{2},-\tfrac{1}{2}\rangle^{\rm II}\\
\vert \tfrac{3}{2},-\tfrac{3}{2}\rangle^{\rm II}
\end{array}
\right.,
\hspace*{0.1cm}
\vert {\rm FI}^+_0\rangle^2=\left\{
\begin{array}{l}
\vert \tfrac{1}{2},\tfrac{1}{2}\rangle^{\rm II}\\
\vert \tfrac{3}{2},\tfrac{1}{2}\rangle^{\rm II}\\
\vert \tfrac{3}{2},\tfrac{3}{2}\rangle^{\rm II}\\
\vert \tfrac{1}{2},-\tfrac{1}{2}\rangle^{\rm II}\\
\vert \tfrac{3}{2},-\tfrac{1}{2}\rangle^{\rm II}\\
\vert \tfrac{3}{2},-\tfrac{3}{2}\rangle^{\rm II}
\end{array}
\right.,
\hspace*{0.15cm}
\vert {\rm FI}^+_0\rangle^3=\left\{
\begin{array}{l}
\vert \tfrac{1}{2},\tfrac{1}{2}\rangle^{\rm II}\\
\vert \tfrac{1}{2},-\tfrac{1}{2}\rangle^{\rm II}
\end{array}
\right.,
\vert {\rm FI}^+_0\rangle^4=\left\{
\begin{array}{l}
\vert \tfrac{1}{2},\tfrac{1}{2}\rangle^{\rm I}\\
\vert \tfrac{1}{2},\tfrac{1}{2}\rangle^{\rm II}\\
\vert \tfrac{1}{2},-\tfrac{1}{2}\rangle^{\rm I}\\
\vert \tfrac{1}{2},-\tfrac{1}{2}\rangle^{\rm II}
\end{array}
\right.,
\nonumber\\
\vert {\rm FI}_0\rangle^5&=\left\{
\begin{array}{l}
\vert \tfrac{1}{2},\tfrac{1}{2}\rangle^{\rm I}\\
\vert \tfrac{1}{2},-\tfrac{1}{2}\rangle^{\rm I}
\end{array}
\right..
\end{align}
At $J<0$:
\begin{align}
\vert {\rm FI}^-_0\rangle^1&=\left\{
\begin{array}{l}
\vert \tfrac{5}{2},\tfrac{5}{2}\rangle\\
\vert \tfrac{5}{2},\tfrac{3}{2}\rangle\\
\vert \tfrac{5}{2},\tfrac{1}{2}\rangle\\
\vert \tfrac{5}{2},-\tfrac{5}{2}\rangle\\
\vert \tfrac{5}{2},-\tfrac{3}{2}\rangle\\
\vert \tfrac{5}{2},-\tfrac{1}{2}\rangle
\end{array}
\right.,
\hspace*{0.1cm}
\vert {\rm FI}^-_0\rangle^2=\left\{
\begin{array}{l}
\vert \tfrac{5}{2},\tfrac{5}{2}\rangle,\\
\vert \tfrac{5}{2},\tfrac{3}{2}\rangle\\
\vert \tfrac{5}{2},\tfrac{1}{2}\rangle\\
\vert \tfrac{3}{2},\tfrac{1}{2}\rangle^{\rm I}\\
\vert \tfrac{3}{2},\tfrac{3}{2}\rangle^{\rm I}\\
\vert \tfrac{5}{2},-\tfrac{5}{2}\rangle,\\
\vert \tfrac{5}{2},-\tfrac{3}{2}\rangle\\
\vert \tfrac{5}{2},-\tfrac{1}{2}\rangle\\
\vert \tfrac{3}{2},-\tfrac{1}{2}\rangle^{\rm I}\\
\vert \tfrac{3}{2},-\tfrac{3}{2}\rangle^{\rm I}
\end{array}
\right.,
\vert {\rm FI}^-_0\rangle^3=\left\{
\begin{array}{l}
\vert \tfrac{3}{2},\tfrac{1}{2}\rangle^{\rm I}\\
\vert \tfrac{3}{2},\tfrac{3}{2}\rangle^{\rm I}\\
\vert \tfrac{3}{2},-\tfrac{1}{2}\rangle^{\rm I}\\
\vert \tfrac{3}{2},-\tfrac{3}{2}\rangle^{\rm I}
\end{array}
\right.,
\vert {\rm FI}^-_0\rangle^4=\left\{
\begin{array}{l}
\vert \tfrac{1}{2},\tfrac{1}{2}\rangle^{\rm I}\\
\vert \tfrac{3}{2},\tfrac{1}{2}\rangle^{\rm I}\\
\vert \tfrac{3}{2},\tfrac{3}{2}\rangle^{\rm I}\\
\vert \tfrac{1}{2},-\tfrac{1}{2}\rangle^{\rm I}\\
\vert \tfrac{3}{2},-\tfrac{1}{2}\rangle^{\rm I}\\
\vert \tfrac{3}{2},-\tfrac{3}{2}\rangle^{\rm I}
\end{array}
\right.,
\nonumber\\
\vert {\rm FI}_0\rangle^5&=\left\{
\begin{array}{l}
\vert \tfrac{1}{2},\tfrac{1}{2}\rangle^{\rm I}\\
\vert \tfrac{1}{2},-\!\tfrac{1}{2}\rangle^{\rm I}
\end{array}
\right..
\end{align}
\subsubsection{\label{reduced bipartite negativities}Reduced bipartite negativities ${\cal N}_{\mu|S_1}$ and ${\cal N}_{S_1|S_2}$}
\begin{figure}[t!]
{\includegraphics[width=0.42\textwidth,trim=3.4cm 9cm 5.85cm 8.5cm, clip]{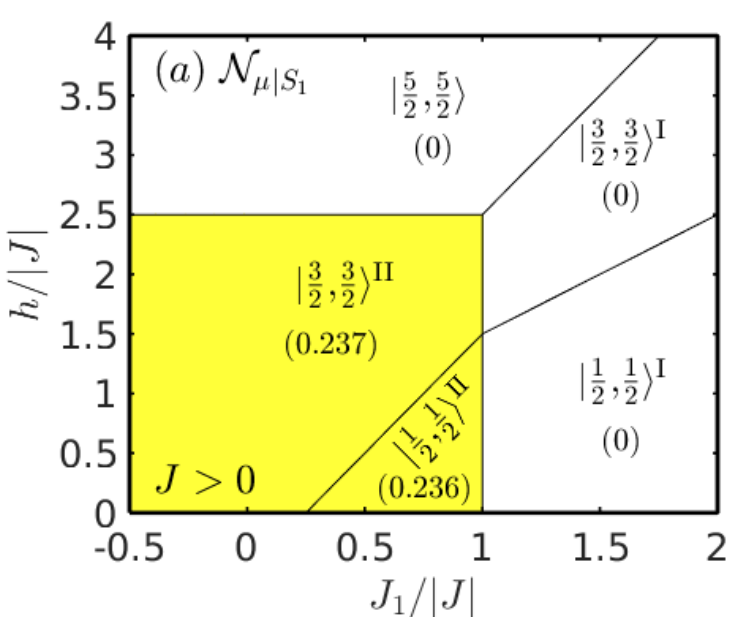}}
{\includegraphics[width=0.47\textwidth,trim=4.35cm 9cm 3.2cm 8.5cm, clip]{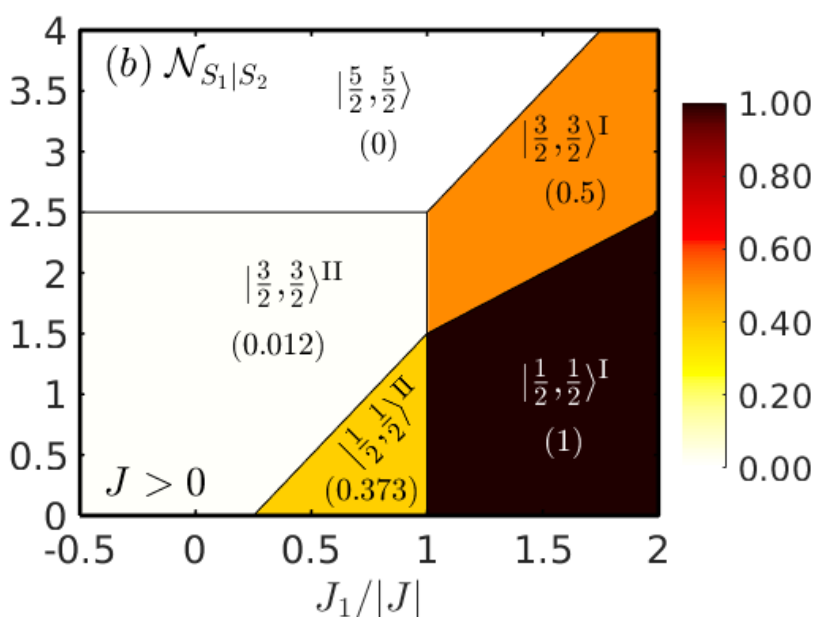}}
\\
{\includegraphics[width=0.42\textwidth,trim=3.4cm 9cm 5.85cm 8.5cm, clip]{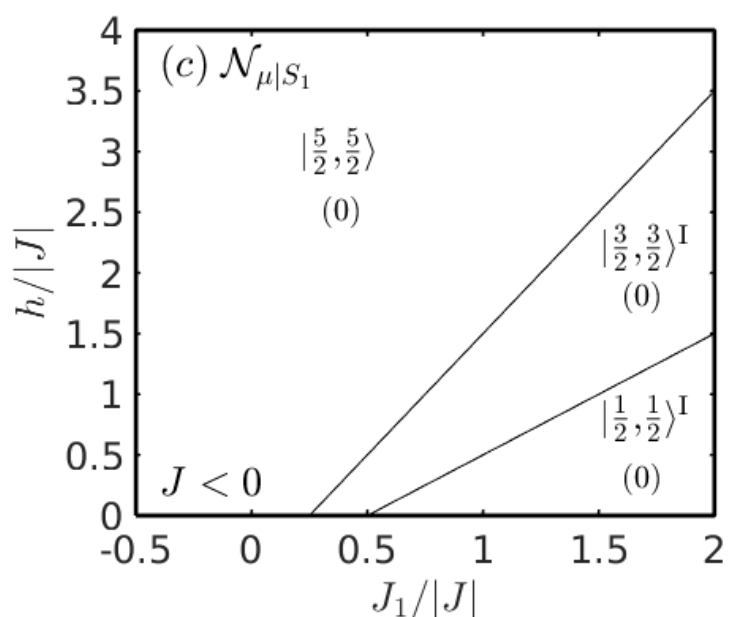}}
{\includegraphics[width=0.47\textwidth,trim=4.35cm 9cm 3.2cm 8.5cm, clip]{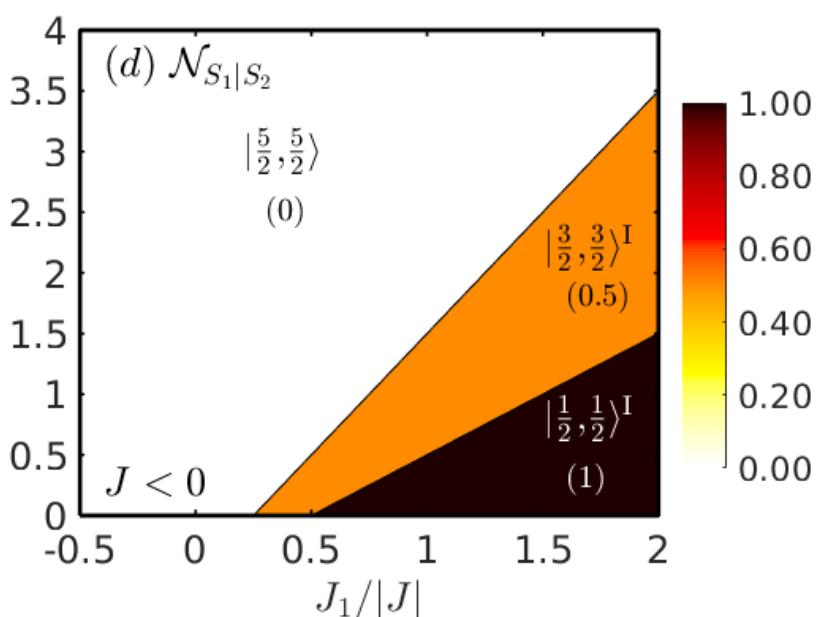}}
\caption{Reduced quantum bipartite negativities,  ${\cal N}_{\mu|S_1}$ and  ${\cal N}_{S_1| S_2}$, of the mixed spin-($1$,$1/2$,$1$) Heisenberg trimer in the $J_1/|J|-h/|J|$ plane. Solid black lines indicate the boundaries between different magnetic ground states.
Numbers in parentheses denote the magnitudes of the reduced bipartite negativity of non-degenerate pure ground states. 
}
\label{fig1}
\end{figure}

The behavior of both reduced bipartite negativities at zero temperature is illustrated in Fig.~\ref{fig1}, alongside the ground-state phase diagram. 
The upper panels illustrate the behavior of the trimer with antiferromagnetic coupling ($J > 0$), while the lower panels depict the behavior with ferromagnetic coupling ($J < 0$). The strength of the reduced bipartite negativity for each non-degenerate pure ground state is indicated in round brackets. The values of the reduced bipartite negativities for all remaining degenerate ground states are summarized in Tab.\ref{tab3}. 
It turns out that reduced bipartite negativity may either emerge simultaneously within each spin dimer or appear solely within the spin-$1$ dimer. The simultaneous presence of non-zero reduced bipartite negativities occurs exclusively in systems with antiferromagnetic exchange interaction ($J > 0$) and when $J_1/J \leq 1$.
Such a state cannot occur spontaneously in the absence of a magnetic field, as shown in Tab.~\ref{tab3}. On the other hand, when the magnetic field is sufficiently strong ($h/|J| > 2.5$), all magnetic moments become aligned along the field direction, leading to a complete suppression of reduced bipartite entanglement in both the spin-$1$ and mixed-spin dimers.

The reduced bipartite negativity confined solely within the spin-$1$ dimer can reach its maximum value \linebreak (${\cal N}_{S_1|S_2} (\vert \tfrac{1}{2},\tfrac{1}{2}\rangle^{\rm I})=1$) or be reduced to one-half  (${\cal N}_{S_1|S_2}(\vert \tfrac{3}{2},\tfrac{3}{2}\rangle^{\rm I})=0.5$)  in two dominant pure non-degenerate ground states. The corresponding reduced density operators, $\hat{\rho}_{S_1|S_2}$, exhibit the structure of maximally or half-entangled spin-$1$ dimers~\cite{Ghannadan2021}. Consequently, no entanglement is expected to be shared between the spin-$1/2$ and spin-$1$ particles in these states.

\begin{table*}[t!]
\caption{Magnitudes of reduced bipartite negativities  ${\cal N}_{S_1|S_2}$, ${\cal N}_{\mu|S_1}$,  as well as the full bipartite negativities ${\cal N}_{S_1|\mu S_2}$,  ${\cal N}_{\mu|S_1S_2}$, along with graphical representations of the reduced bipartite entanglements ${\cal N}_{A|B}$ contributing to the global tripartite negativity  ${\cal N}_{\mu S_1S_2}$ at degenerate ground states. In the triangle diagram, the top vertex represents spin $\mu$, while the other two correspond to spins $S_1$, $S_2$. Thick (thin) edges indicate non-zero (zero) values of ${\cal N}_{A|B}$. The last column classifies the type of tripartite entanglement (TE) according to  Sab\'{i}n and Garc\'{i}a-Alcaine~\cite{Sabin}.}
\label{tab3}
\centering
\begin{tabular}{c| l |cc |c c | c c | c}
&ground state & ${\cal N}_{S_1|S_2}$ &${\cal N}_{\mu|S_1}$&    ${\cal N}_{S_1|\mu S_2}$ &   ${\cal N}_{\mu|S_1S_2}$&   ${\cal N}_{\mu S_1S_2}$& ${\cal N}_{A|B}$ & type of TE~\cite{Sabin}
 \\
\hline\hline
\multicolumn{6}{l}{   $J_1/|J|<1:$
}  \\
  \hline
\multicolumn{1}{ c|  }{\multirow{2}{*}{ \multirow{2}{*}{$\xuparrow{0.4cm}$ $h/|J|$, $J>0$ }} }                     
&  $\vert {\rm FI}\rangle^{\rm II}_{\tfrac{5}{2}-\tfrac{3}{2}}$ &0.003&0.034&0.042   &  0.070  & 0.049 &
 \begin{tikzpicture}
\draw[very thick] (0,0)  -- (1/3,0)  --(1/6,1/3)-- cycle;
\draw (1/6,1/3) circle [radius=0.07];
\end{tikzpicture}   
 &  {\it subtype 2-3}
  \\
 \multicolumn{1}{ c|  }{}                        
&  $\vert {\rm FI}\rangle^{\rm II}_{\tfrac{3}{2}-\tfrac{1}{2}}$ &0.046&0.095&0.388   &  0.436  & 0.403 &
 \begin{tikzpicture}
\draw[very thick] (0,0)  -- (1/3,0)  --(1/6,1/3)-- cycle;
\draw (1/6,1/3) circle [radius=0.07];
\end{tikzpicture}   
 &  {\it subtype 2-3}
  \\
  \hline
 \multicolumn{6}{l}{   $J_1/|J|>1:$} \\
  \hline
\multicolumn{1}{ c|  }{\multirow{2}{*}{ $\xuparrow{0.4cm}$ $h/|J|$\hspace*{1cm}} }                        
&  $\vert {\rm FI}\rangle^{\rm I}_{\tfrac{5}{2}-\tfrac{3}{2}}$ &0.104&0&0.104   &  0  & 0 & 
 \begin{tikzpicture}
\draw[] (0,0)  -- (1/3,0)  --(1/6,1/3)-- cycle;
\draw[very thick] (0,0)  -- (1/3,0) -- cycle;
\draw (1/6,1/3) circle [radius=0.07];
\end{tikzpicture}   
 & {\it subtype 1-1}
  \\
 \multicolumn{1}{ c|  }{}                        
&  $\vert {\rm FI}\rangle^{\rm I}_{\tfrac{3}{2}-\tfrac{1}{2}}$ &0.423&0&0.423   &  0  & 0 &
 \begin{tikzpicture}
\draw[] (0,0)  -- (1/3,0)  --(1/6,1/3)-- cycle;
\draw[very thick] (0,0)  -- (1/3,0) -- cycle;
\draw (1/6,1/3) circle [radius=0.07];
\end{tikzpicture}   
 &  {\it subtype 1-1}
 \\
\hline\hline
\multicolumn{6}{l}{    $J_1/|J|=1:$} \\
  \hline
\multicolumn{1}{ c|  }{\multirow{5}{*}{ $\xuparrow{1.2cm}$ $h/|J|$, $J>0$} } &
$\vert {\rm FI}_1\rangle^{2}$ &0.028&0.016&0.067    &  0.047   & 0.059   &
 \begin{tikzpicture}
\draw[very thick] (0,0)  -- (1/3,0)  --(1/6,1/3)-- cycle;
\draw (1/6,1/3) circle [radius=0.07];
\end{tikzpicture}   
&  {\it subtype 2-3}
  \\
  \multicolumn{1}{ c|  }{}                        &
  $\vert {\rm FI}\rangle_{\tfrac{3}{2}-\tfrac{3}{2}}$ &0.083& 0.056& 0.245   &  0.200    & 0.229&
 \begin{tikzpicture}
\draw[very thick] (0,0)  -- (1/3,0)  --(1/6,1/3)-- cycle;
\draw (1/6,1/3) circle [radius=0.07];
\end{tikzpicture}   
& {\it subtype 2-3}
  \\
  \multicolumn{1}{ c|  }{}                        &
  $\vert {\rm FI}_1\rangle^{1}$ &0.115& 0.014&0.244   &  0.147 & 0.206&
 \begin{tikzpicture}
\draw[very thick] (0,0)  -- (1/3,0)  --(1/6,1/3)-- cycle;
\draw (1/6,1/3) circle [radius=0.07];
\end{tikzpicture}   
  &   {\it subtype 2-3}
 \\
 \multicolumn{1}{ c|  }{}                        &
  $\vert {\rm FI}\rangle_{\tfrac{1}{2}-\tfrac{1}{2}}$ &0.346& 0.061& 0.702   &  0.236    & 0.488 &
 \begin{tikzpicture}
\draw[very thick] (0,0)  -- (1/3,0)  --(1/6,1/3)-- cycle;
\draw (1/6,1/3) circle [radius=0.07];
\end{tikzpicture}   
&  {\it subtype 2-3}
 \\
 \multicolumn{1}{ c|  }{}                        &
  $\vert {\rm FI}_0^+\rangle^{4}$ &0.25&0&0.5   &  0.167  & 0.347 & 
 \begin{tikzpicture}
\draw[] (0,0)  -- (1/3,0)  --(1/6,1/3)-- cycle;
\draw[very thick] (0,0)  -- (1/3,0) -- cycle;
\draw (1/6,1/3) circle [radius=0.07];
\end{tikzpicture}   
 & {\it subtype 2-1}
  \\
  \hline\hline
\multicolumn{6}{l}{               $h/|J|=0:$}
  \\
  \hline
\multicolumn{1}{ c|  }{\multirow{5}{*}{ $\xuparrow{1.2cm}$ $J_1/|J|$, $J>0$} } &
$\vert {\rm FI_0}\rangle^{5}$ &1&0& 1   &  0  &0   & 
 \begin{tikzpicture}
\draw[] (0,0)  -- (1/3,0)  --(1/6,1/3)-- cycle;
\draw[very thick] (0,0)  -- (1/3,0) -- cycle;
\draw (1/6,1/3) circle [radius=0.07];
\end{tikzpicture}   
&{\it subtype 1-1}
\\
 \multicolumn{1}{ c|  }{}                        &
  $\vert {\rm FI_0^+}\rangle^{4}$ &0.25&0&0.5 &  0.167 &0.347   &
 \begin{tikzpicture}
\draw[] (0,0)  -- (1/3,0)  --(1/6,1/3)-- cycle;
\draw[very thick] (0,0)  -- (1/3,0) -- cycle;
\draw (1/6,1/3) circle [radius=0.07];
\end{tikzpicture}   
 &  {\it subtype 2-1} 
 \\
    \multicolumn{1}{ c|  }{}        &
  $\vert {\rm FI_0^+}\rangle^{3}$ &0.333&0&0.839  &  0.333 & 0.617     & 
 \begin{tikzpicture}
\draw[] (0,0)  -- (1/3,0)  --(1/6,1/3)-- cycle;
\draw[very thick] (0,0)  --(1/3,0);
\draw (1/6,1/3) circle [radius=0.07];
\end{tikzpicture}   
& {\it subtype 2-1}
   \\
  \multicolumn{1}{ c|  }{}        &
 $\vert {\rm FI_0^+}\rangle^{2}$ &0&0.111&0.184  &  0.245 & 0.202  &
 \begin{tikzpicture}
\draw[] (0,0)  -- (1/3,0)  --(1/6,1/3)-- cycle;
\draw[very thick] (0,0)  -- (1/6,1/3)  --(1/3,0);
\draw (1/6,1/3) circle [radius=0.07];
\end{tikzpicture}   
&   {\it subtype 2-2}
\\
  \multicolumn{1}{ c|  }{}        &
 $\vert {\rm FI_0^+}\rangle^{1}$ &0&0.166&0.174 &  0.200 &0.182   &
 \begin{tikzpicture}
 \draw[] (0,0)  -- (1/3,0)  --(1/6,1/3)-- cycle;
\draw[very thick] (0,0)  -- (1/6,1/3)  --(1/3,0);
\draw (1/6,1/3) circle [radius=0.07];
\end{tikzpicture}   
 &  {\it subtype 2-2}
  \\
 \hline
\multicolumn{1}{ c|  }{\multirow{5}{*}{ $\xuparrow{1.2cm}$ $J_1/|J|$, $J<0$} } &
$\vert {\rm FI_0}\rangle^{5}$ &1&0& 1   &  0  &0   & 
 \begin{tikzpicture}
\draw[] (0,0)  -- (1/3,0)  --(1/6,1/3)-- cycle;
\draw[very thick] (0,0)  -- (1/3,0) -- cycle;
\draw (1/6,1/3) circle [radius=0.07];
\end{tikzpicture}   
&{\it subtype 1-1}
\\
 \multicolumn{1}{ c|  }{}                        &
  $\vert {\rm FI_0^-}\rangle^{4}$ &0.111&0&0.263 &0  &0  &
 \begin{tikzpicture}
\draw[] (0,0)  -- (1/3,0)  --(1/6,1/3)-- cycle;
\draw[very thick] (0,0)  -- (1/3,0) -- cycle;
\draw (1/6,1/3) circle [radius=0.07];
\end{tikzpicture}   
 &  {\it subtype 1-1} 
 \\
    \multicolumn{1}{ c|  }{}        &
  $\vert {\rm FI_0^-}\rangle^{3}$ &0.333&0& 0.399   &0 & 0   & 
 \begin{tikzpicture}
\draw[] (0,0)  -- (1/3,0)  --(1/6,1/3)-- cycle;
\draw[very thick] (0,0)  --(1/3,0);
\draw (1/6,1/3) circle [radius=0.07];
\end{tikzpicture}   
& {\it subtype 1-1}
   \\
  \multicolumn{1}{ c|  }{}        &
 $\vert {\rm FI_0^-}\rangle^{2}$ &0&0&0 &0   &0   &
 \begin{tikzpicture}
\draw[] (0,0)  -- (1/3,0)  --(1/6,1/3)-- cycle;
\draw (1/6,1/3) circle [radius=0.07];
\end{tikzpicture}   
&   {\it subtype 0-0}
\\
  \multicolumn{1}{ c|  }{}        &
 $\vert {\rm FI_0^-}\rangle^{1}$ &0&0&0 &0  & 0  &
 \begin{tikzpicture}
 \draw[] (0,0)  -- (1/3,0)  --(1/6,1/3)-- cycle;
\draw (1/6,1/3) circle [radius=0.07];
\end{tikzpicture}   
 &  {\it subtype 0-0}
  \\
 \hline
 \end{tabular}
\end{table*}

\subsubsection{\label{full bipartite negativity} Full bipartite negativities ${\cal N}_{\mu|S_1S_2}$ and ${\cal N}_{S_1|\mu S_2}$}

\begin{figure}[t!]
{\includegraphics[width=0.455\textwidth,trim=3.2cm 9cm 5.8cm 8.5cm, clip]{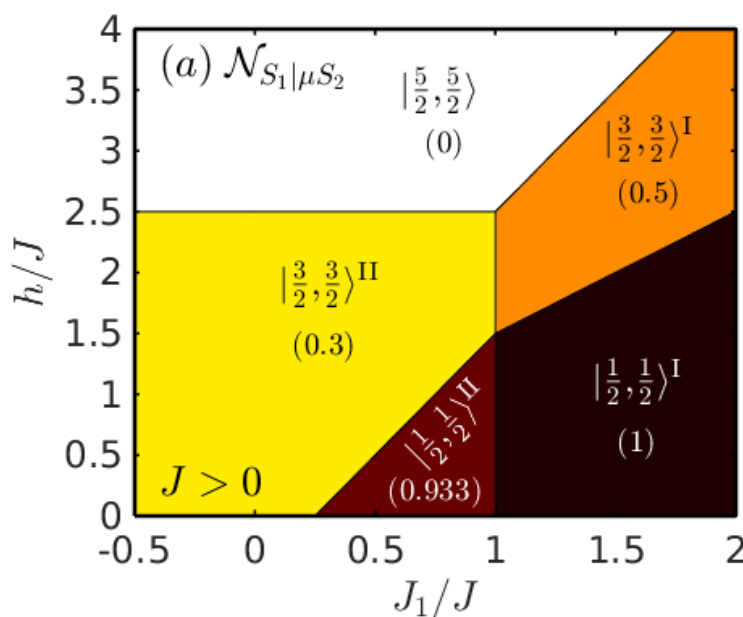}}
{\includegraphics[width=0.53\textwidth,trim=4.1cm 9cm 3.cm 8.5cm,clip]{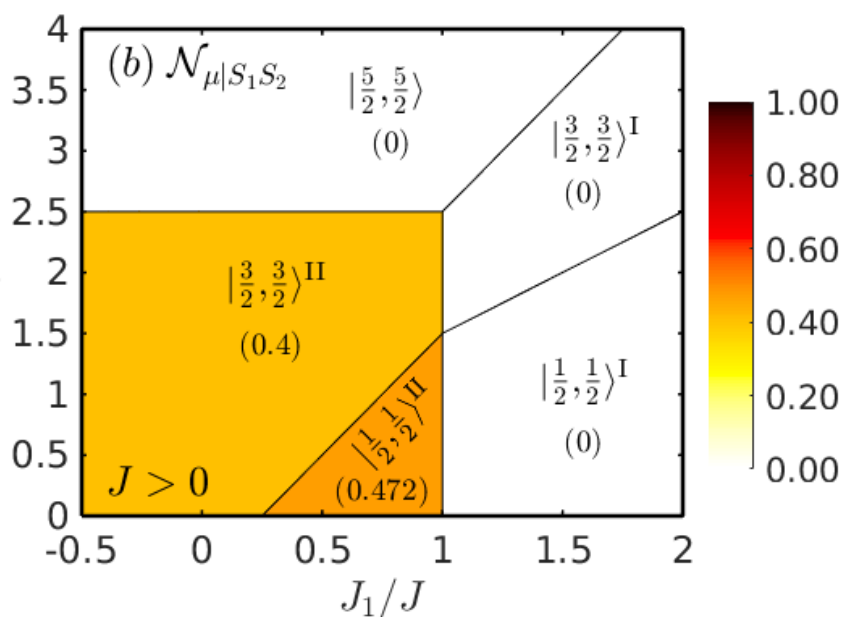}}
\caption{Zero-temperature density plots of bipartite negativity ($a$) ${\cal N}_{S_1|\mu S_2}$ and ($b$) ${\cal N}_{\mu|S_1S_2}$ in the $J_1/J-h/J$ plane. Numbers in parentheses indicate the strengths of bipartite negativities for the corresponding non-degenerate pure ground states.
}
\label{fig2}
\end{figure}

Fig.~\ref{fig2}  illustrates the behavior of the full bipartite negativities ${\cal N}_{S_1|\mu S_2}$ and ${\cal N}_{\mu|S_1S_2}$ for $J > 0$, with partial results that can be generalized to the $J < 0$ case, respecting the corresponding ground-state phase diagram.
In the regions corresponding to the non-degenerate ground states $\vert \tfrac{1}{2},\tfrac{1}{2}\rangle^{\rm I}$ and $\vert \tfrac{3}{2},\tfrac{3}{2}\rangle^{\rm I}$, the magnitude of the full bipartite negativity ${\cal N}_{S_1|\mu S_2}$ is identical to that of its reduced counterpart ${\cal N}_{S_1|S_2}$, indicating that the maximum entanglement is shared exclusively between the two spin-$1$ particles, $S_1$ and $S_2$. Consequently, the full bipartite negativity ${\cal N}_{\mu|S_1S_2}$ vanishes in these regions.
For the non-degenerate ground states $\vert \tfrac{1}{2},\tfrac{1}{2}\rangle^{\rm II}$ and $\vert \tfrac{3}{2},\tfrac{3}{2}\rangle^{\rm II}$, both full bipartite negativities are expected to be non-zero. However, the differing magnitudes of entanglement observed in these states suggest a distinct redistribution of quantum correlations within the spin trimer, depending on the applied magnetic field.

Significantly smaller magnitudes of full bipartite negativities are observed along the boundary lines, where degeneracy plays a crucial role in stabilizing entanglement. The corresponding data are summarized in Tab.~\ref{tab3}. Notably, unexpected behavior is identified for two degenerate ferrimagnetic ground states, $\vert {\rm FI}_0^+\rangle^3$ and $\vert {\rm FI}_0^+\rangle^4$. In both cases, the full bipartite negativity ${\cal N}_{\mu|S_1S_2}$ remains relatively strong, while the reduced bipartite negativity ${\cal N}_{\mu|S_1}$ is zero.

Mathematically, this discrepancy originates from the structure of the partially transposed reduced density matrix $\hat{\rho}_{\mu S_1}^{T_{\mu}}$~\eqref{a2} and, more specifically, from the relatively simple analytical expression for the corresponding eigenvalue~\eqref{a4}:
\begin{align}
(\lambda_2^{\mp})_{\mu S_1}=\tfrac{1}{2}\left\{\omega_{11}^{\mp}+\omega_{22}^{\pm}-\sqrt{(\omega_{11}^{\mp}+\omega_{22}^{\pm})^2-4[\omega_{11}^{\mp}\omega_{22}^{\pm}-(\omega_{24}^{\mp})^2]} \right\}.
\nonumber
\end{align}
This expression requires the term $\left(\omega_{11}^{\mp}\omega_{22}^{\pm}-(\omega_{24}^{\mp})^2\right)$ to be negative. However, in both of the aforementioned ground states, this condition is violated due to degeneracy. In contrast, the two negative eigenvalues of the full transposed density matrix $\hat{\rho}_{\mu S_1S_2}^{T{\mu}}$ are provided in Eq.~\eqref{a13}:
\begin{align}
(\lambda_5^{\mp})_{\mu}&=\left\{\rho_{3,3}^{\mp}-\sqrt{(\rho_{3,3}^{\mp})^2+4(\rho_{3,11}^{\pm})^2]} \right\}, \hspace*{2.5cm}\mbox{for $\vert {\rm FI}_0^+\rangle^3$}\nonumber\\
(\lambda_5^{\mp})_{\mu}&=\tfrac{1}{2}\left\{\rho_{3,3}^{\mp}-\rho_{3,7}^{\mp}-\sqrt{(\rho_{3,3}^{\mp}-\rho_{3,7}^{\mp})^2+16(\rho_{3,11}^{\pm})^2]} \right\}, \mbox{for $\vert {\rm FI}_0^+\rangle^4$}.
\end{align}
Since $\rho_{3,11}^{\pm}$ is non-zero in both cases, these negative eigenvalues lead to a non-zero bipartite negativity within the full tripartite system.

From a physical perspective, this behavior can be interpreted as a collective phenomenon that generates bipartite entanglement between the individual  spin-$1/2$ and the spin-$1$ dimer, which is considered a composite of two non-separable entities rather than two individual spin-$1$ particles. This type of entanglement is particularly fragile in the tripartite system, as tracing out one spin-$1$ completely destroys the entanglement within the spin-$1$ dimer and, consequently, the entanglement between the mixed-spin dimer.

\subsubsection{\label{tripartite} The global tripartite entanglement}

\begin{figure}[t!]
\centering
{\includegraphics[width=0.55\textwidth,trim=3.2cm 8.7cm 3.cm 8.5cm, clip]{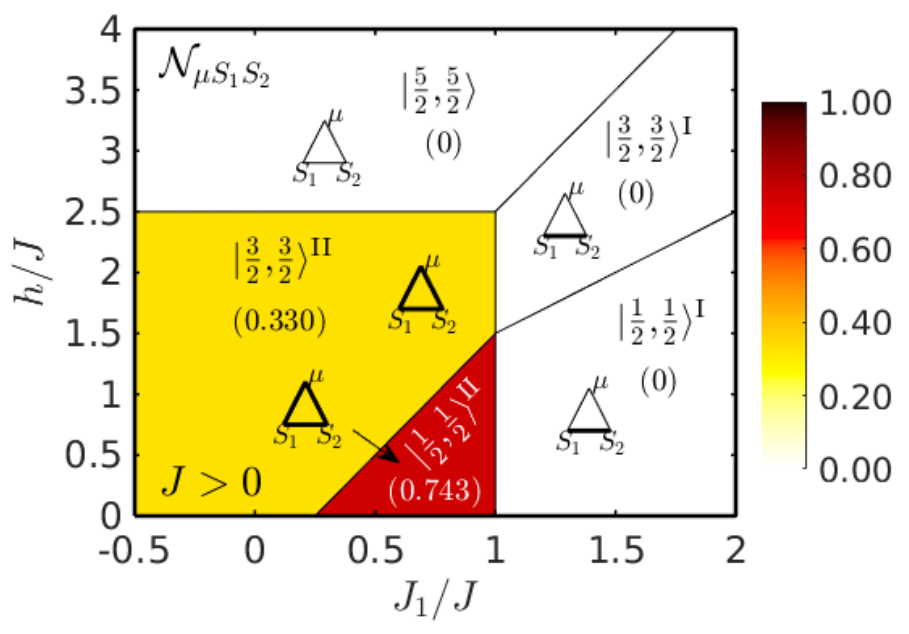}}
\caption{Zero-temperature density plot of  the global tripartite negativity  ${\cal N}_{\mu S_1S_2}$ in the $J_1/J-h/J$ plane. Numbers in parentheses denote the strengths of tripartite negativity for the corresponding non-degenerate pure ground states. The distribution of reduced bipartite negativities is schematically illustrated by the thickness of the triangle edges: thick (thin) edges represent entangled (separable) subsystems.
}
\label{fig3}
\end{figure}

The behavior of the global tripartite negativity is illustrated in Fig.\ref{fig3}, with additional data provided in Tab.\ref{tab3}. We unambiguously identify the presence of global entanglement among all three spins for an antiferromagnetic exchange interaction $J>0$ under the specific coupling ratio $J_1/J\leq 1$,  provided the magnetic field is smaller than  $h/J=2.5$. 
In the case of a non-zero magnetic field,  global tripartite entanglement is accompanied by a bipartite entanglement between each pair of spins.
Following the classification of tripartite entangled states introduced by Sab\'{i}n and Garc\'{i}a-Alcaine~\cite{Sabin}, such states can be identified as {\it  full inseparable states of the subtype 2-3}. This type of tripartite entanglement is robust, as it remains intact under the loss (i.e., tracing out) of any one of the three spins.
 This observation is particularly relevant for future studies of multipartite (global) entanglement, as the theoretically predicted region of the parameter space closely corresponds to that of real trinuclear complexes discussed in the Introduction. Their nearly linear three-spin geometry promotes antiferromagnetic exchange coupling between mixed spins ($J$) while significantly suppressing the exchange interaction between identical spins, i.e., $J_1/J \to 0$.

In the absence of a magnetic field, we observe weaker but non-negligible global tripartite negativity in all realized ground states, as shown in Tab.~\ref{tab3}. For sufficiently low exchange interaction ($J_1/J \leq 1/4$), tripartite entanglement is mediated through the spin-$1/2$ particle, with non-zero reduced bipartite negativity appearing in the mixed-spin dimers. These states are classified as {\it  full inseparable states of the subtype 2-2}~\cite{Sabin}. As the exchange interaction increases ($1/4 < J_1/J \leq 1$), the reduced bipartite entanglement shifts toward the two identical spin-$1$ particles, while the system remains globally entangled. This case is referred to as a {\it  full inseparable states of the subtype 2-1}~\cite{Sabin}.  
Interestingly, in {\it subtype 2-1}, the spin-$1/2$ particle is not entangled with either spin-$1$ in any reduced subsystem, which might seem counterintuitive. Nevertheless, the global negativity for {\it subtype 2-1} is significantly greater than that of {\it subtype 2-2}, owing to stronger correlations between the two spin-$1$ particles.

Outside this parameter space, the tripartite negativity vanishes, implying that the system is either fully separable or biseparable. Since the Hilbert space dimension of the model exceeds $2 \otimes 3$, zero negativity (by the PPT criterion) is a necessary but not sufficient condition for separability~\cite{Peres,Horodecki}. Thus, we cannot definitively determine the separability of these states.
To remain consistent with the Garc\'{i}a-Alcaine classification, we categorize all ground states with zero tripartite negativity based on the presence or absence of reduced bipartite negativities ${\cal N}_{\mu|S_1}$ and ${\cal N}_{S_1|S_2}$.
We identify the  ferromagnetic ground state $\vert \tfrac{5}{2},\tfrac{5}{2}\rangle$ as a fully separable state of  {\it type 0-0} showing no bipartite entanglement. Two pure non-degenerate states $\vert \tfrac{3}{2},\tfrac{3}{2}\rangle^{\rm I}$ and $\vert \tfrac{1}{2},\tfrac{1}{2}\rangle^{\rm I}$  are classified as biseparable states of  {\it subtype 1-1}. In these states, the loss (or tracing over) of the separable spin-$1/2$ still leaves entanglement between the two spin-$1$ particles. The degenerate states along the boundary lines, $\vert {\rm FI}\rangle^{\rm I}_{\tfrac{5}{2}-\tfrac{3}{2}}$ and $\vert {\rm FI}\rangle^{\rm I}_{\tfrac{3}{2}-\tfrac{1}{2}}$,  also fall into  {\it subtype 1-1}. 
Other ground states with zero tripartite negativity that are stable in zero magnetic field are either fully separable or biseparable. States $\vert {\rm FI}_0^-\rangle^{1}$ and $\vert {\rm FI}_0^-\rangle^{2}$, which exhibit neither full nor reduced bipartite entanglement, are classified as fully separable of {\it type 0-0}. 
States $\vert {\rm FI}_0^-\rangle^{3}$, $\vert {\rm FI}_0^-\rangle^{4}$, and $\vert {\rm FI}0\rangle^{5}$ show separability between the spin-$1/2$ and the spin-$1$ dimer but retain non-zero entanglement between the spin-$1$ particles (${ \cal N}_{S_1|S_2} \ne 0$). These states are categorized as biseparable states of {\it subtype 1-1}~\cite{Sabin}.

\subsection{\label{Thermal stability} Thermal stability of tripartite entanglement}
\begin{figure}[t!]
{\includegraphics[width=0.34\textwidth,trim=3.cm 8.5cm 5.55cm 8.5cm,clip]{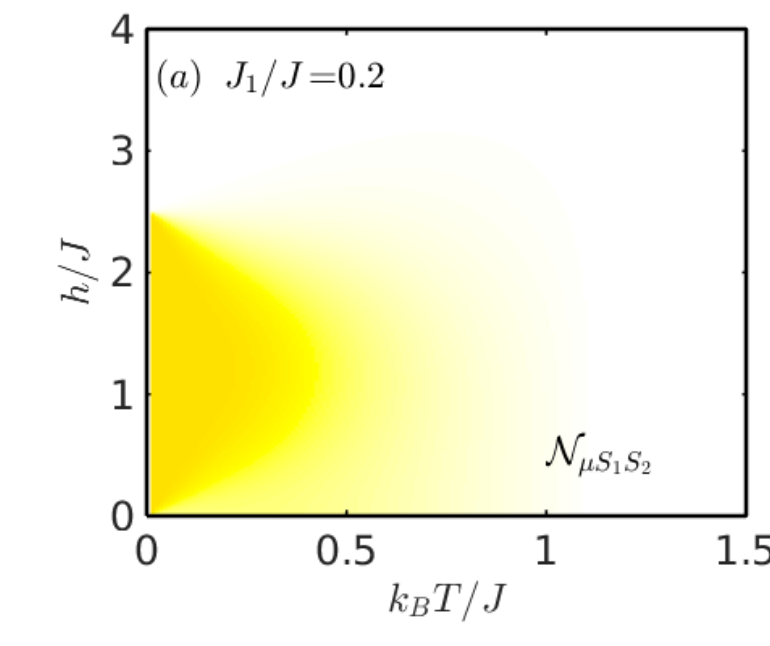}}
{\includegraphics[width=0.29\textwidth,trim=4.9cm 8.5cm 5.55cm 8.5cm, clip]{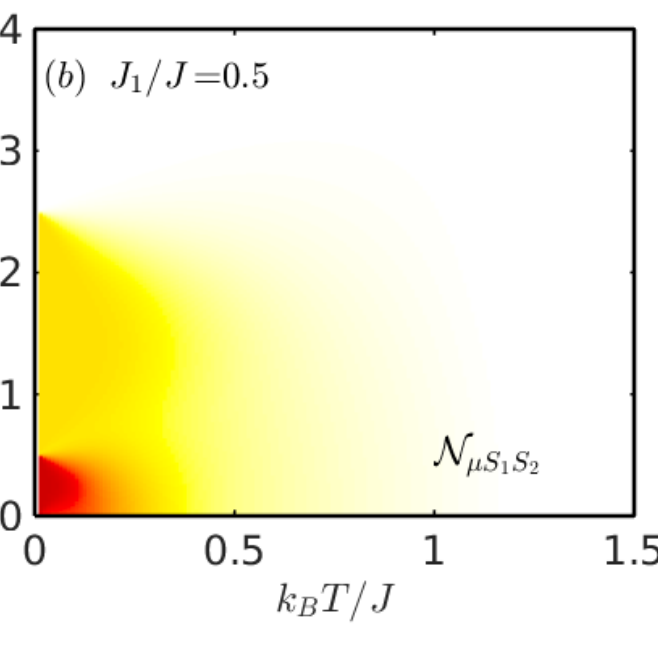}}
{\includegraphics[width=0.35\textwidth,trim=4.9cm 8.5cm 3.2cm 8.5cm,clip]{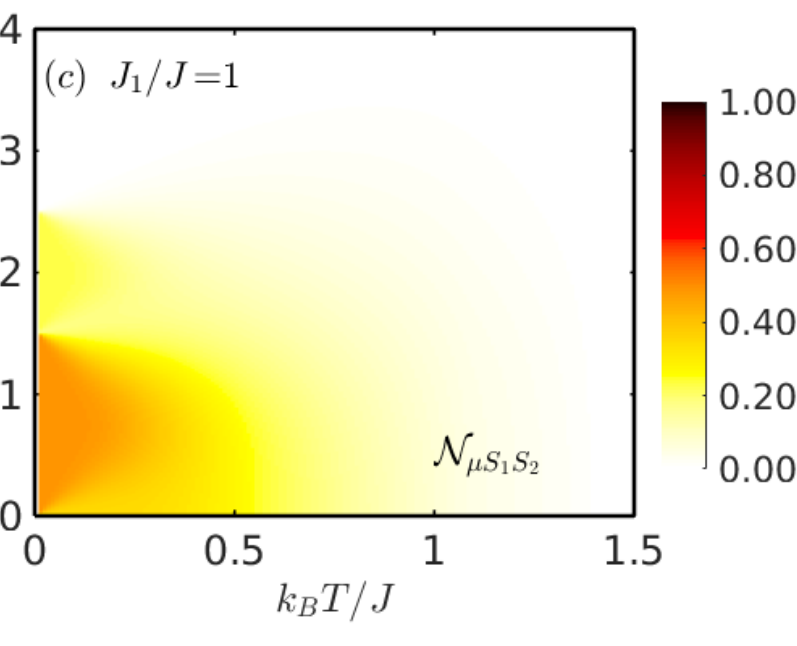}}
\caption{Density plots of  global tripartite entanglement $\mu S_1S_2$ in the $k_BT/J$-$h/J$ plane for three representative values of the exchange ratio  $J_1/J\leq1$, with $J>0$. 
}
\label{fig4}
\end{figure}

Let us now examine the thermal stability of the detected tripartite entangled states determined at zero temperature. The typical behavior for three representative ratios of $J_1/J$ ($J>0$) is illustrated in Fig.~\ref{fig4}.
The tripartite negativity decreases gradually with increasing temperature, reflecting the underlying characteristics of the respective ground-state entanglements.  As the exchange coupling ratio  $J_1/J$  increases, a slight enhancement in the threshold temperature (marking the limit beyond which tripartite entanglement vanishes) is observed. 

\begin{figure}[b!]
\centering
{\includegraphics[width=0.5\textwidth,trim=3.cm 8.5cm 2.5cm 8.5cm,clip]{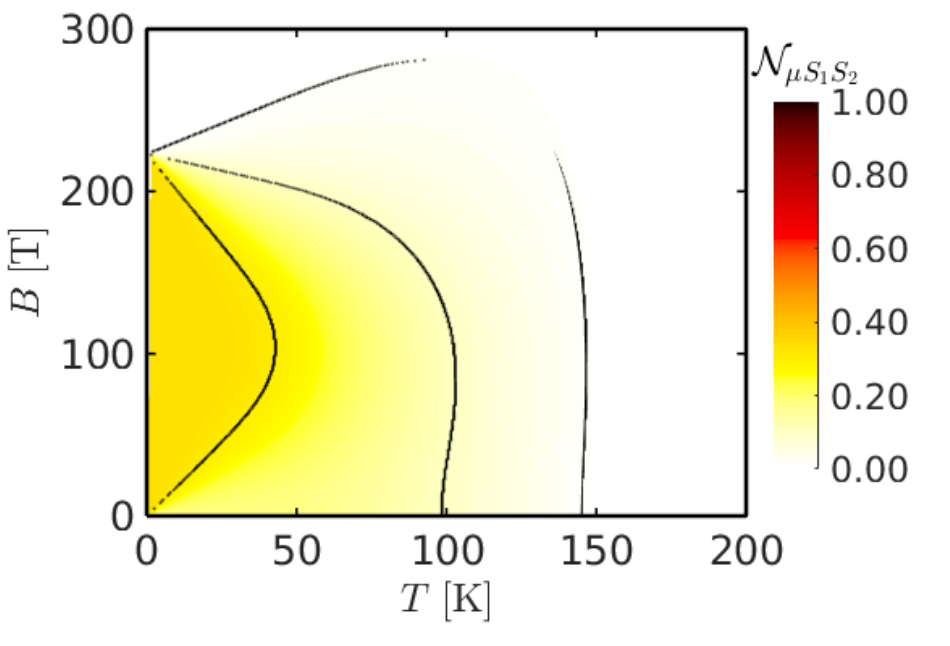}}
\caption{A density plot of global tripartite negativity ${\cal N}_{\mu S_1S_2}$ in the temperature-field plane conforming to the Ni$_2$Cu compound, with model parameters  $J/k_B=90.3$ cm$^{-1}$, $J_1/k_B=0$ cm$^{-1}$, and the weighted gyromagnetic factor $g=(2g_{\rm Ni}\!+\!g_{\rm Cu})/2$, adapted from Ref.~\cite{Ribas}. Black contour lines indicate negativity values of $0.3$, $0.1$, and $0.01$ (from left to right). }
\label{fig4a}
\end{figure}

 To connect our theoretical analysis with the behavior of real materials, we next employ a set of experimentally determined parameters to predict the thermal robustness of tripartite entanglement in a three-spin complex, [Ni(bapa)(H$_2$0)]$_2$Cu(pba)(ClO$_4$)$_2$,  where bapa stands for bis($3$-aminopropyl)amine, and pba denotes $1$,$3$-propylenebis(oxamato)~\cite{Ribas}. 
Experimental measurements on the Ni$_2$Cu complex revealed a relatively strong antiferromagnetic exchange coupling of $J/k_B=90.3$ cm$^{-1}$ between the Ni-Cu ions, and a negligible exchange interaction between the two Ni ions, with $J_1/k_B\approx 0$ cm$^{-1}$. Additionally, two distinct but closely spaced  Land\'e $g$ factors were determined: $g_{Ni}=2.19$ and $g_{Cu}=2.12$. For simplicity, our theoretical analysis uses their weighted average,  $g=(2g_{Ni}\!+\!g_{Cu})/3\doteq 2.1667$. 
The qualitative behavior of thermal tripartite entanglement (Fig.~\ref{fig4a}) closely resembles that shown in Fig.~\ref{fig4}($a$), as expected in the limiting case $J_1/J\to 0$. Surprisingly, the quantitative analysis indicates a relatively high resistance to thermal fluctuations, likely due to the strong antiferromagnetic exchange coupling $J/k_B$. A global tripartite entanglement of sufficient intensity, ${\cal N}\gtrsim 0.1$, persists up to temperatures of approximately $T\lesssim 100$ K and magnetic fields $B\lesssim 210$ T. Moreover, a weak thermal tripartite entanglement remains detectable at temperatures exceeding $150$ K.  Although the observed threshold temperature remains below room temperature, it is nearly three times higher than that reported for the opposite spin configuration, consisting of a central spin $S=1$ and two peripheral spins $\mu_i=1/2$~\cite{Ghannadan2025}. This notable enhancement suggests that quantum systems with a higher total spin of magnetic molecule may exhibit improved thermal robustness of entanglement. Consequently, complex spin architectures could offer a promising pathway for extending operational temperature ranges in future quantum technologies.
\begin{figure}[t!]
{\includegraphics[width=0.32\textwidth,trim=4.cm 8.5cm 3.1cm 8.5cm,clip]{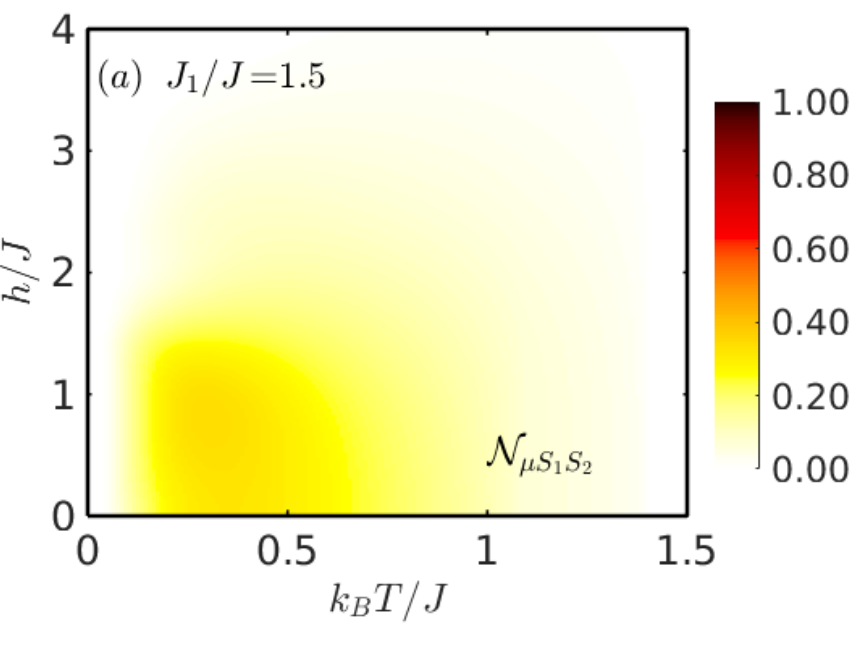}}{\includegraphics[width=0.34\textwidth,trim=3.1cm 8.4cm 3.1cm 9.1cm, clip]{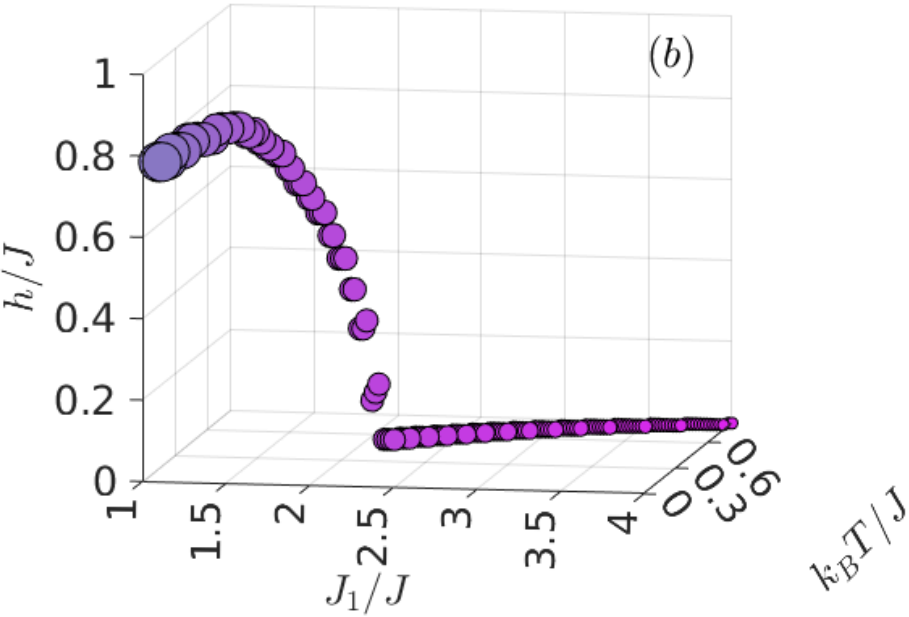}}
{\includegraphics[width=0.32\textwidth,trim=3.cm 8.4cm 4cm 8.5cm, clip]{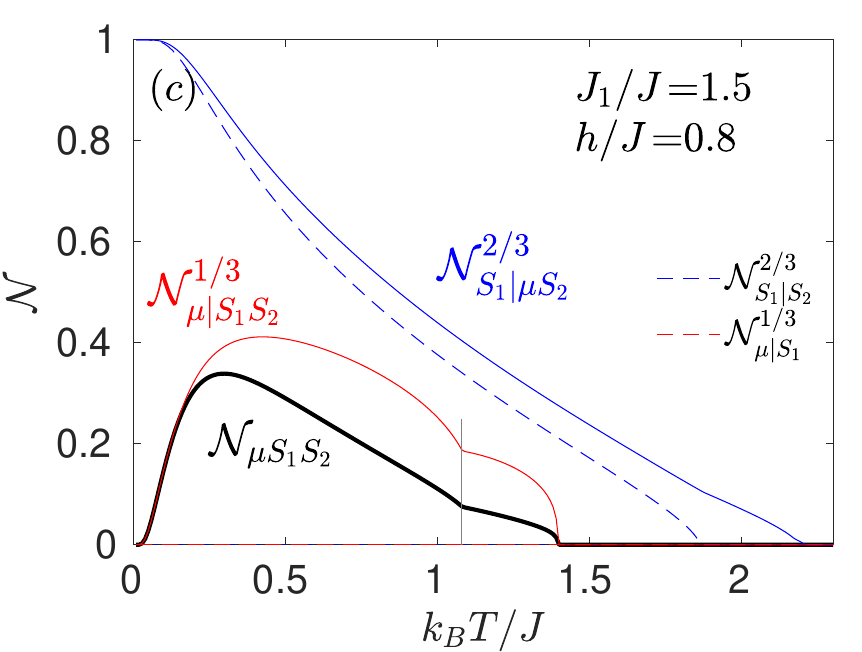}}
\caption{($a$) Density plot of   tripartite entanglement ${\cal N}_{\mu S_1S_2}$ in the $k_BT/J$-$h/J$ plane for  the ratio $J_1/J=1.5$.  ($b$) The movement of the maximum thermal  tripartite entanglement  ${\cal N}^{max}_{\mu S_1S_2}$ in the $J_1/J$-$k_BT/J$-$h/J$ space. The magnitude of ${\cal N}^{max}_{\mu S_1S_2}$  is indicated by the size and color of the circles (both radius and darkness increase with greater negativity), ranging from approximately $\sim 0.488$ 
 to $\sim 0.139$. ($c$) Thermal  tripartite negativity ${\cal N}_{\mu S_1S_2}$ (black solid line) for $J_1/J=1.5$ and $h/J=0.8$, together with its bipartite contributions,  ${\cal N}^{1/3}_{\mu|S_1S_2}$ (red solid line), ${\cal N}^{1/3}_{\mu |S_1}$ (red dashed line), ${\cal N}^{2/3}_{S_1|\mu S_2}$ (blue solid line), and ${\cal N}^{2/3}_{S_1|S_2}$ (blue dashed line).}
\label{fig5}
\end{figure}
An additional intriguing observation from our analysis arises in the region of the $\vert \tfrac{1}{2},\tfrac{1}{2}\rangle^{\rm I}$ ground state with antiferromagnetic $J$ coupling. At zero temperature, this pure state is biseparable, with its density operator factorized into the tensor product of two subsystems, with a maximally entangled spin-$1$ dimer, $\vert \tfrac{1}{2},\tfrac{1}{2}\rangle^{\rm I}\langle \tfrac{1}{2},\tfrac{1}{2}\vert^{\rm I}=\vert \tfrac{1}{2}\rangle\langle\tfrac{1}{2}\vert\otimes[\vert1,\!-\!1\rangle-\vert 0,0\rangle+\vert \!-\!1,1\rangle]\otimes[\langle1,\!-\!1\vert-\langle 0,0\vert+\langle \!-\!1,1\vert]$. 
Consequently, the tripartite entanglement at zero temperature is zero. As shown in Fig.~\ref{fig5}($a$), increasing the temperature unexpectedly activates  tripartite entanglement of substantial magnitude. While slight thermal activation of entanglement (on the order of about $0.01$) is a well-known phenomenon attributed to the thermal population of low-lying excited states, in this case the maximum tripartite negativity reaches a relatively high value of approximately  $0.338$. The magnitude of ${\cal N}_{\mu S_1S_2}$ varies with $J_1/J$, $h/J$, and $k_BT/J$, ranging from $0.488$ to $0.139$ within the region   $J_1/J\in(1,4)$. This variation is illustrated in Fig.~\ref{fig5}($b$), where the size and color intensity of circles represent negativity, where larger and darker circles indicate higher values. We observe that the maximum strength of entanglement decreases with increasing  $J_1/J$, primarily due to the rise in temperature. Furthermore, for $J_1/J\lesssim 2$, the tripartite negativity shows a non-trivial increase with increasing magnetic field, maintaining its maximum value at non-zero $h/J$.

The activation of tripartite negativity can be attributed to the redistribution of entanglement from the maximally entangled spin-$1$ dimer to the combined system comprising the spin-$1$ dimer and the individual spin-$1/2$ particle. Specifically, the progressive population of higher-energy states gradually reduces the negativity ${\cal N}_{S_1|\mu S_2}$, which is primarily governed by the entanglement within the spin-$1$ dimer, ${\cal N}_{S_1|S_2}$. For illustration, see Fig.~\ref{fig5}($c$), which presents representative values of $J_1/J=1.5$ and $h/J=0.8$. Simultaneously, a pronounced increase in the ${\cal N}_{\mu|S_1S_2}$ contribution is observed. This enhancement results from the activation of quantum correlations between the spin-$1$ dimer and the individual spin-$1/2$ particle, while strong quantum correlations within the spin-$1$ dimer persist (as indicated by the dashed blue line in  Fig.~\ref{fig5}($c$)). As a consequence, the thermally activated, relatively large tripartite negativity reflects a global entanglement shared simultaneously among all three spins. We hypothesize that this non-trivial and  substantial degree of  ${\cal N}_{\mu S_1S_2}$  emerges if and only if the partially separable ground state exhibits a sufficiently strong entanglement within one of the system’s subsystems (e.g., ${\cal N}_{S_1|S_2}=1$).
\begin{figure}[t!]
{\includegraphics[width=0.45\textwidth,trim=3.cm 8.4cm 4.1cm 9.05cm, clip]{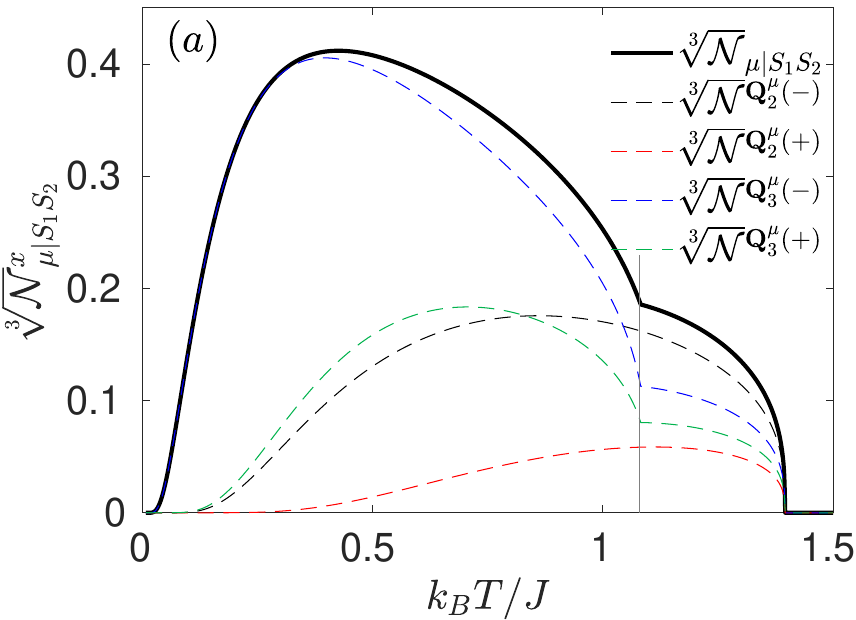}}
{\includegraphics[width=0.5\textwidth,trim=3.cm 8.4cm 2.5cm 9.05cm, clip]{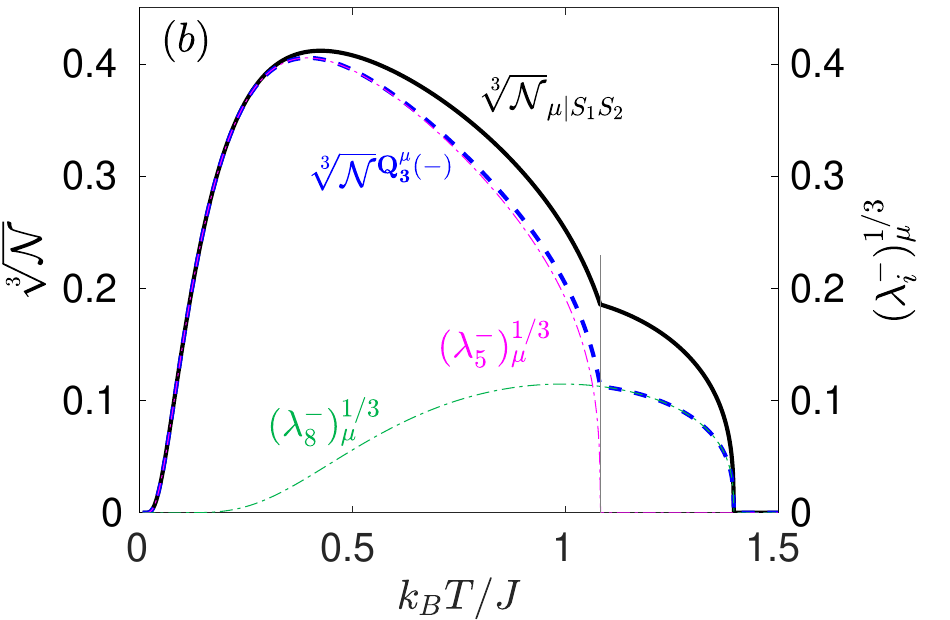}}
\caption{($a$) Thermal bipartite negativity $\sqrt[3]{\cal N}_{\mu|S_1S_2}$ (black solid line) for $J_1/J=1.5$ and $h/J=0.8$, along with its partial contributions originating from the matrices $\mathbf{Q}^{\mu}_2(-)$ (black dashed line), $\mathbf{Q}^{\mu}_2(+)$ (red dashed line), $\mathbf{Q}^{\mu}_3(-)$ (blue dashed line), and $\mathbf{Q}^{\mu}_3(+)$ (green dashed line).
($b$) Thermal behavior of the negative eigenvalues $(\lambda_5^{-})_{\mu}^{1/3}$ and $(\lambda_8^{-})_{\mu}^{1/3}$, shown together with the bipartite negativity $\sqrt[3]{\cal N}_{\mu|S_1S_2}$ (black solid line) and its
 $\mathbf{Q}^{\mu}_3(-)$ contribution (blue dashed line). The gray vertical line indicates the location of the singular point resulting from the sign change of the $(\lambda_5^-)_{\mu}$ eigenvalue.  }
\label{fig6}
\end{figure}

To gain a deeper understanding of the mechanisms behind the thermal activation of negativity  ${\cal N}_{\mu |S_1S_2}$,  we perform a more detailed analysis of the partially transposed density matrix $\hat{\rho}_{\mu S_1S_2}^{T_{\mu}}$, as defined in the Appendix.  The  matrix $\hat{\rho}_{\mu S_1S_2}^{T_{\mu}}$ exhibits a block-diagonal structure composed of six distinct blocks $\mathbf{Q}^{\mu}_1(\mp)$, $\mathbf{Q}^{\mu}_2(\mp)$ and $\mathbf{Q}^{\mu}_3(\mp)$, Eq.~\eqref{a12}. While the eigenvalues of the blocks $\mathbf{Q}^{\mu}_1(\mp)$ are always positive, the bipartite negativity ${\cal N}_{\mu|S_1S_2}$ arises solely from the negative eigenvalues $(\lambda_3^{\mp})_{\mu}$, $(\lambda_5^{\mp})_{\mu}$, and $(\lambda_8^{\mp})_{\mu}$ of the remaining four blocks.
 As shown in Fig.~\ref{fig6}($a$), the low-temperature behavior of  ${\cal N}_{\mu|S_1S_2}$ contribution is primarily governed by the negative eigenvalues of the matrix
$\mathbf{Q}^{\mu}_3(-)$ (blue dashed line). Specifically, by the $(\lambda_5^{-})_{\mu}$ eigenvalue (Eq.~\eqref{a13}) as illustrated in Fig.~\ref{fig6}($b$). This behavior becomes prominent when the condition
\begin{align}
(\lambda_5^{-})_{\mu}:\;\;& (\rho_{2,2}^{-}-\rho_{2,4}^{-})(\rho_{3,3}^{+}-\rho_{3,7}^{+})<(\rho_{3,11}^{-}-\rho_{3,13}^{-})^2,\label{eq8c}
\end{align}
is satisfied, which with the help of Eq.~\eqref{a9} leads to the  relation $\beta J>\tfrac{2}{3}\ln 4$. Thus, thermal activation of global entanglement in the $\vert \tfrac{1}{2},\tfrac{1}{2}\rangle^{\rm I}$ ground state is possible
only for antiferromagnetic exchange interaction $J>0$. Furthermore, the structure of the $(\lambda_5^-)_{\mu}$ eigenvalue specifies the relevance of five energy states $\varepsilon^{\rm II}_{\tfrac{1}{2},\mp\tfrac{1}{2}}$, $\varepsilon^{\rm I}_{\tfrac{3}{2},\mp\tfrac{1}{2}}$, and $\varepsilon^{\rm I}_{\tfrac{3}{2},\tfrac{3}{2}}$. Among these, the first excited state $\varepsilon^{\rm II}_{\tfrac{1}{2},\tfrac{1}{2}}$, which exhibits strong bipartite entanglement both in ${\cal N}_{\mu|S_1S_2}$ and  ${\cal N}_{S_1|\mu S_2}$, is responsible for the rapid thermal enhancement of  negativity in the region $J_1/J>1$.

Moreover, Fig.~\ref{fig6}($b$) clearly reveals the origin of an interesting singular point observed in the ${\cal N}_{\mu|S_1S_2}$ curve, indicated by the vertical gray line. At this point, the eigenvalue $(\lambda_5^{-})_{\mu}$  ceases to contribute to the negativity due to the sign change from negative to positive. Consequently, only the negative contribution from $(\lambda_8^{-})_{\mu}$ governs the character of the $\mathbf{Q}^{\mu}_3(-)$ block in the bipartite negativity ${\cal N}_{\mu|S_1S_2}$.

\subsection{\label{theory vs. experiment}Theoretical study of entanglement vs. its experimental determination}
Finally, we would like to make a few notes regarding the connection between the theoretically predicted behavior of entanglement and its experimental measurement. Although negativity is not a directly measurable quantity, its definition,  Eq.~\eqref{eq2}, is intrinsically linked to the density matrix $\hat{\rho}$ of a  microscopic model that accurately describes the magnetic properties of the investigated system. Naturally, one potential approach to estimating the negativity is to reconstruct the density matrix via experimentally accessible observables. Given that the elements of the density matrix are directly related to various local observables through the relation, $\langle \hat{\cal O}\rangle=\mathrm{Tr} \left(\hat{\cal O}\hat{\rho}\right)$, our goal is to identify a complete set of local expectation values that uniquely determine the density matrix $\hat{\rho}$.

It is straightforward to show that five local observables, $\langle \hat{\mu}^z\rangle$, $\langle \hat{S}^z_i\rangle$, 
$\langle \hat{\mu}^z\hat{S}^z_i\rangle$, $\langle \hat{S}_1^z\hat{S}^z_2\rangle$, and $\langle (\hat{S}_1^z)^2\hat{S}^z_2\rangle$, can be expressed in terms of the diagonal elements $\rho_{i,i}^{\mp}$ of the full system's density matrix (see Eq.~\eqref{a9}):
\begin{align}
\langle \hat{\mu}^z\rangle&=\mathrm{Tr} \left(\hat{\rho}\hat{\mu}^z\right)=\frac{1}{2}\left\{
 (\rho_{1,1}^{-}+2\rho_{2,2}^{-}+2\rho_{3,3}^{-}+
  \rho_{5,5}^{-}+2\rho_{6,6}^{+}+\rho_{9,9}^{+})\right.
  \nonumber\\
  &\left.\hspace*{2.3cm}-(\rho_{1,1}^{+}+2\rho_{2,2}^{+}+2\rho_{3,3}^{+}+
  \rho_{5,5}^{+}+2\rho_{6,6}^{-}+\rho_{9,9}^{-} ) \right\},
  \nonumber\\
\langle \hat{S}^z_i\rangle&=\mathrm{Tr} \left(\hat{\rho}\hat{S}_i^z\right)=
(\rho_{1,1}^{-}+\rho_{2,2}^{-}+ \rho_{6,6}^{-}+ \rho_{9,9}^{-})-(\rho_{1,1}^{+}+ \rho_{2,2}^{+}+\rho_{6,6}^{+}  +\rho_{9,9}^{+}),
   \nonumber\\
\langle \hat{\mu}^z\hat{S}_i^z\rangle&=\mathrm{Tr} \left(\hat{\rho}\hat{\mu}^z\hat{S}^z_i\right)=\frac{1}{2}\left\{
(\rho_{1,1}^{-}+ \rho_{2,2}^{-}- \rho_{6,6}^{-}- \rho_{9,9}^{-})+ (\rho_{1,1}^{+} + \rho_{2,2}^{+}- \rho_{6,6}^{+}-\rho_{9,9}^{+})\right\},
  \nonumber\\
\langle \hat{S}_1^z\hat{S}_2^z\rangle&=\mathrm{Tr} \left(\hat{\rho}\hat{S}_1^z\hat{S}^z_2\right)=
 (\rho_{1,1}^{-}-2\rho_{3,3}^{-}+\rho_{9,9}^{-})+(\rho_{1,1}^{+} -2\rho_{3,3}^{+}+\rho_{9,9}^{+}),
  \nonumber\\  
\langle (\hat{S}_1^z)^2\hat{S}^z_2\rangle&=\mathrm{Tr}\left( \hat{\rho}(\hat{S}_1^z)^2\hat{S}^z_2\right)=(\rho_{1,1}^{-}+\rho_{9,9}^{-})-(\rho_{1,1}^{+}+\rho_{9,9}^{+}).
\end{align}
After straightforward calculations, one can derive  explicit analytical expressions of all diagonal elements of the density matrix as functions of sublattice magnetizations, $\langle \mu^z\rangle$, $\langle S_i^z\rangle$, along with the  local spin correlation functions $\langle \mu^zS^z_i\rangle$, $\langle S_1^zS^z_2\rangle$,  and $\langle (S_1^z)^2S^z_2\rangle$. During this derivation, the normalization condition ${\rm Tr} \hat{\rho}=\sum_{i=1}^9 [\rho_{i,i}^{-}+\rho_{i,i}^{+}]=1$ must also be taken into account. Thus,
\begin{align}
\rho_{1,1}^{-}\!&=\!\frac{\alpha_5}{E_1}, \;\; \;\;
\nonumber\\
\rho_{2,2}^{-}\!&=\!-2{\rm e}^{-\beta h}\cosh(\beta h)\rho_{1,1}^{-}\!+\!\frac{\alpha_3}{C_1}\!+\!\frac{\alpha_4}{2D_1},\;\;\;\;
\nonumber\\
\rho_{3,3}^{-}\!&=\!{\rm e}^{-2\beta h}\cosh(2\beta h)\rho_{1,1}^{-}\!-\!\frac{\langle S_1^zS_2^z\rangle{\rm e}^{\frac{\beta h}{2}}}{4\cosh\left(\frac{\beta h}{2}\right)}\!+\!{\rm e}^{-\beta h}[2\cosh(\beta h)-1]\frac{\alpha_4}{2D_1},
\nonumber\\
\rho_{5,5}^{-}\!&=\!\frac{{\rm e}^{\frac{\beta}{2}h}}{2\cosh\left(\frac{\beta h}{2}\right)}\!-\!{\rm e}^{-2\beta h}\left[4\cosh^2(\beta h)\!-\!2\cosh(\beta h)\!-\!1\right]\rho_{1,1}^{-}\!-\!2[\rho_{3,3}^{-}\!+\!\rho_{6,6}^{+}]
\!-\!{\rm e}^{-\beta h}\left[2\cosh(\beta h)\!-\!1\right]\left[\rho_{9,9}^{+}\!+\!2\rho_{2,2}^{-}\right],
\nonumber\\
\rho_{9,9}^{-}\!&=\!{\rm e}^{-\beta h}\rho_{1,1}^{-}\!+\!\frac{\alpha_4}{D_1},
\nonumber\\
\rho_{6,6}^{-}\!&=\!-2{\rm e}^{-2\beta h}\cosh(\beta h)\rho_{1,1}^{-}\!+\!\frac{\alpha_1}{A_1}\!-\!2\frac{\alpha_4}{D_1}{\rm e}^{-\beta h}\sinh^2\left(\frac{\beta h}{2}\right)\!+\!{\rm e}^{-\beta h}\frac{\alpha_3}{C_1}.
\end{align}
Of course, the following relations hold between the corresponding diagonal elements of the density matrix: $\rho^+_{1,1}={\rm e}^{-5\beta h}\rho^-_{1,1}$, $\rho^+_{2,2}={\rm e}^{-3\beta h}\rho^-_{2,2}$, $\rho^+_{3,3}={\rm e}^{-\beta h}\rho^-_{3,3}$, $\rho^+_{5,5}={\rm e}^{-\beta h}\rho^-_{5,5}$, $\rho^+_{6,6}={\rm e}^{-\beta h}\rho^-_{6,6}$, and $\rho^+_{9,9}={\rm e}^{-3\beta h}\rho^-_{9,9}$. Due to the complexity of the resulting expressions, we introduce the following abbreviations:
\begin{align}
A_1\!&=\!-8{\rm e}^{-\beta h}\sinh(\beta h),\;\;\;\;
C_1\!=\!-\frac{A_1^2}{2}{\rm e}^{-\frac{\beta h}{2}}\sinh\left(\frac{\beta h}{2}\right),\;\;\;\;
D_1\!=\!A_1C_1{\rm e}^{-\beta h}\cosh^2\left(\frac{\beta h}{2}\right),\;\;\;\;\nonumber\\
E_1\!&=\!-A_1D_1{\rm e}^{-\frac{3}{2}\beta h}\cosh(\beta h)\cosh\left(\frac{\beta h}{2}\right),\;\;\;\;\nonumber
\end{align}
\begin{align}
\alpha_1\!&=\!2{\rm e}^{-\frac{\beta}{2}h}\left\{ 2\langle \mu^z\rangle\cosh\left(\frac{\beta h}{2}\right)\!-\!\sinh\left(\frac{\beta h}{2}\right)\right\},\nonumber\\
\alpha_2\!&=\!-2{\rm e}^{-\frac{\beta}{2}h}\left\{ \langle S_1^z\rangle\cosh\left(\frac{\beta h}{2}\right)\!+\!2\langle\mu^zS_1^z\rangle\sinh\left(\frac{\beta h}{2}\right)\right\},\nonumber\\
\alpha_3\!&=\!2{\rm e}^{-\frac{\beta}{2} h}\left\{ \alpha_1\cosh\left(\frac{\beta h}{2}\right)\!-\!8\langle\mu^zS_1^z\rangle{\rm e}^{-\frac{\beta h}{2}}\sinh(\beta h)\right\},\nonumber\\
\alpha_4\!&=\!-\frac{A_1}{2}{\rm e}^{-\beta h}\left\{ 
2\alpha_3\cosh(\beta h)\!+\!\alpha_2A_1{\rm e}^{\frac{\beta}{2} h}\sinh\left(\frac{\beta h}{2}\right)\right\},\nonumber\\
\alpha_5\!&=\!-2{\rm e}^{-\frac{3}{2}\beta h}\sinh\left(\frac{\beta h}{2} \right)[2\cosh(\beta h)+1]\alpha_4\!+\!D_1\langle (S_1^z)^2S_2^z\rangle.
\end{align}
The off-diagonal elements  $\rho_{i,j}^{\mp}$,  can be determined through the following four local expectation values $\langle \hat{\mu}^x\hat{S}^x_i\rangle$, $\langle \hat{S}_1^x\hat{S}^x_2\rangle$, $\langle \hat{\mu}^x(\hat{S}_1^x)^2\hat{S}^x_2\rangle$, and $\langle (\hat{S}_1^x)^2(\hat{S}^x_2)^2\rangle$
\begin{align}
\langle \hat{\mu}^x\hat{S}^x_i\rangle&=\mathrm{Tr}\; \hat{\rho}\hat{\mu}^x\hat{S}^x_i=\frac{1}{\sqrt{2}}\left[
 (\rho_{2,10}^{-}+\rho_{3,11}^{-}+\rho_{5,11}^{-})+
 (\rho_{2,10}^{+}+\rho_{3,11}^{+}+\rho_{5,11}^{+})\right],
  \nonumber\\
\langle \hat{S}_1^x\hat{S}^x_2\rangle&=\mathrm{Tr}\; \hat{\rho}\hat{S}_1^x\hat{S}_2^x=
 (\rho_{2,4}^{-}+2\rho_{3,5}^{-}+\rho_{6,8}^{-})+
 (\rho_{2,4}^{+}+2\rho_{3,5}^{+}+\rho_{6,8}^{+}),
  \nonumber\\
\langle \hat{\mu}^x(\hat{S}_1^x)^2\hat{S}^x_2\rangle&=\mathrm{Tr}\; \hat{\rho}\hat{\mu}^x(\hat{S}_1^x)^2\hat{S}^x_2=\frac{1}{2\sqrt{2}}
\left[  
(\rho_{2,10}^{-}+\rho_{3,11}^{-}+\rho_{3,13}^{-}+2\rho_{5,11}^{-})+
(\rho_{2,10}^{+}+\rho_{3,11}^{+}+\rho_{3,13}^{+}+2\rho_{5,11}^{+})\right],
  \nonumber\\
\langle (\hat{S}_1^x)^2(\hat{S}^x_2)^2\rangle&=\mathrm{Tr}\; \hat{\rho}(\hat{S}_1^x)^2(\hat{S}^x_2)^2=\frac{1}{4}
\left[ (\rho_{1,1}^{-}+4\rho_{2,2}^{-}+2\rho_{3,3}^{-}+4\rho_{5,5}^{-}+ 4\rho_{6,6}^{-}+\rho_{9,9}^{-}+2\rho_{3,7}^{-})\right.
\nonumber\\
&\left.\hspace*{3.3cm}+ (\rho_{1,1}^{+}+ 4\rho_{2,2}^{+}+2\rho_{3,3}^{+}+4\rho_{5,5}^{+}+ 4\rho_{6,6}^{+}+\rho_{9,9}^{+}+2\rho_{3,7}^{+})\right].
\end{align}
Bearing in mind the additional identities:
\begin{align}
&\rho_{2,4}^{-}\!=\!{\rm e}^{\beta h}\left(\frac{}{}\rho_{3,7}^{-}\!+\!\rho_{3,5}^{-}\!+\!\frac{1}{\sqrt{2}}\rho_{3,13}^{-}\right),
\hspace*{3cm}
\rho_{3,11}^{-}=\rho_{2,10}^{-}{\rm e}^{-\beta h}-\rho_{3,13}^{-},\nonumber\\
&\rho_{5,11}^{-}=\rho_{2,10}^{-}{\rm e}^{-\beta h},
\hspace*{5.5cm}
\rho_{6,8}^{-}=\rho_{2,4}^{-}{\rm e}^{-\beta h}-\sqrt{2}\rho_{3,13}^{-}.
\end{align}
we identify that
\begin{align}
&\rho_{2,10}^{-}\!=\!\frac{\sqrt{2}{\rm e}^{\tfrac{3}{2}\beta h}}{4\cosh^3(\tfrac{\beta h}{2})}\langle \mu^x(S^x_1)^2S^x_2\rangle,
\hspace*{1.5cm}
\rho_{3,13}^{-}\!=\!{\rm e}^{-\beta h}
\left[2\cosh(\beta h)\!+\!1\right]\rho_{2,10}^{-}\!-\!\frac{\sqrt{2}{\rm e}^{\tfrac{\beta}{2} h}}{2\cosh(\tfrac{\beta h}{2})}\langle\mu^xS^x_1\rangle,\nonumber\\
&\rho_{3,5}^{-}\!=\!\frac{1}{\sqrt{2}}\rho_{3,13}^{-}\!+\!
\frac{{\rm e}^{\tfrac{\beta}{2} h}}{4\cosh(\tfrac{\beta h}{2})}\langle S_1^xS_2^x\rangle,\nonumber\\
&\rho_{3,7}^{-}\!=\!\frac{{\rm e}^{-2\beta h}}{2}\left\{ [4\cosh^2(\beta h)\!-\!2\cosh(\beta h)\!-\!1]\rho_{1,1}^{-}\!+\![2\cosh(\beta h)\!-\!1][4\rho_{2,2}^{-}\!+\!\rho_{9,9}^{+}]\right.\left.\!+\!2{\rm e}^{2\beta h}[\rho_{3,3}^{-}\!+\!2\rho_{5,5}^{-}\!+\!2\rho_{6,6}^{+}]\right\}
\nonumber\\
&\hspace*{1.5cm}-2\langle (S_1^x)^2(S_2^x)^2\rangle.
\end{align}
The remaining off-diagonal elements satisfy the following relations:
 $\rho^+_{2,4}={\rm e}^{-3\beta h}\rho^-_{2,4}$, $\rho^+_{2,10}={\rm e}^{-3\beta h}\rho^-_{2,10}$, $\rho^+_{3,5}={\rm e}^{-\beta h}\rho^-_{3,5}$, $\rho^+_{3,7}={\rm e}^{-\beta h}\rho^-_{3,7}$, $\rho^+_{3,11}={\rm e}^{-\beta h}\rho^-_{3,11}$, $\rho^+_{3,13}={\rm e}^{-\beta h}\rho^-_{3,13}$, $\rho^+_{5,11}={\rm e}^{-\beta h}\rho^-_{5,11}$, and $\rho^+_{6,8}={\rm e}^{-\beta h}\rho^-_{6,8}$. 

In the above, we have clearly demonstrated that the density matrix of the mixed spin-($1$,$1/2$,$1$) Heisenberg trimer~\eqref{eq1} can be  constructed from experimentally accessible quantities. Some of these,  such as $\langle \mu^z\rangle$ and $\langle S^z_i\rangle$, can be directly obtained via standard magnetometry experiments. Meanwhile, other quantities, such as local spin correlation functions, can be indirectly extracted from relevant structure factors measured by inelastic neutron scattering experiments~\cite{Haraldsen,Cruz, Nagler, Su}.   
It should be noted, however, that this procedure is non-trivial, as the relationships  between the density matrix elements and specific local observables depend sensitively on the theoretical model employed.

\section{\label{conclusion}Conclusion}
In this paper, we have analytically investigated the global tripartite entanglement in the mixed spin-($1$,$1/2$,$1$) Heisenberg trimer. The entanglement has been quantified using negativity, derived from the negative eigenvalues of a partially transposed density matrix. The global tripartite negativity has been determined as the geometric mean of all bipartite contributions, and the type of tripartite entanglement has been classified according to the categorization proposed by Sab\'{i}n and Garc\'{i}a-Alcaine~\cite{Sabin}. 
It has been shown that the analyzed system exhibits global entanglement among all three spins in the case of ferrimagnetic ground states, provided that the antiferromagnetic exchange coupling within the mixed-spin dimer $J$ is greater than or equal to the exchange coupling within the spin-$1$ dimer  $J_1$. The observed global tripartite entanglement is distributed over all three spin dimers, with non-zero values for all reduced negativities, when a non-zero external magnetic field is applied. Conversely, in the absence of an external magnetic field, global tripartite entanglement is characterized by one reduced negativity being zero, occurring either within the subsystem of identical spins($J_1/J\leq1/4$) or different  spins ($1/4<J_1/J\leq1$).

Furthermore, we have demonstrated that thermal fluctuations gradually suppress  global tripartite entanglement up to a threshold temperature, beyond which the global tripartite entanglement is completely destroyed. Increasing the strength of the $J_1$ exchange coupling enhances the system’s resistance to thermal fluctuations, resulting in a slight increase in the threshold temperature. We further have predicted the thermal stability of the real three-spin complex Ni$_2$Cu, for which thermally robust global entanglement was found despite the nearly linear arrangement of magnetic ions in the molecule. In this system, global entanglement was predicted to persist up to approximately $100$ K and magnetic fields approaching $210$ T.
In addition, we have identified a remarkable and unexpected  phenomenon whereby thermal fluctuations induce global tripartite entanglement of  unusually strong magnitude within a region characterized by a biseparable ground state. While  slight thermal induction of global entanglement on the order of $0.01$ is a known effect, to the best of our knowledge, such a pronounced enhancement has not been previously reported. The maximum  global tripartite negativity in this region has been observed near the isotropic case $J_1/J=1$, reaching a value of approximately $0.488$. Further increases in the exchange ratio  $J_1/J$ lead to a modest reduction of this maximum, which nonetheless remains significant at higher ratios, e.g., ${\cal N}_{\mu S_1S_2}(J_1/J=4)\sim 0.139$. This non-trivial thermal entanglement arises from modifications in the entanglement between the central spin $\mu$ and the spin-$1$ dimer, linked to the population of five specific energy states  in the relevant partially transposed density matrix. This behavior is characteristic of systems with exclusively antiferromagnetic exchange interactions $J$.

Finally, we have outlined a possible approach to connect theoretical predictions of entanglement with experimental observations. We have derived that each density matrix element can be expressed in terms of specific local expectation values. Some of these can be directly measured using standard magnetometry experiments, while others can be inferred indirectly from inelastic neutron scattering experiments, which probe structure factors containing information on local spin correlations.
\\\\
{\bf Acknowledgments}\\
\newline
I would like to express my sincere gratitude to Jozef Stre\v{c}ka for his insightful comments and valuable discussions. This work was supported by the Slovak Research and Development Agency under contract No. APVV-20-0150 and by the Slovak Academy of Sciences through grant No. VEGA 1/0298/25.
\\
\appendix
{\bf Appendix}\\
\section{\label{App A} }

\setcounter{equation}{0}
\setcounter{table}{0}
\renewcommand{\theequation}{\thesection.\arabic{equation}}
\renewcommand{\thetable}{\thesection.\arabic{table}}

The analytic expression of the partition function, denoted as ${\cal Z}=\sum_{i=1}^{18} {\rm e}^{-\beta \varepsilon_i}$, is subsequently expressed as follows
\begin{align}
{\cal Z}&=2{\rm e}^{-\beta J_1}\big\{
\cosh\left(\tfrac{5}{2}\beta h\right){\rm e}^{-\beta J}
+
\cosh\left(\tfrac{3}{2}\beta h\right)\left[
2{\rm e}^{\tfrac{\beta}{4} J}\cosh\left(\tfrac{5}{4}\beta J\right)+{\rm e}^{-\tfrac{\beta}{2}(J-4J_1)}\right]
\nonumber\\
&+
\cosh\left(\tfrac{\beta}{2}h \right)\left[
2{\rm e}^{\tfrac{\beta}{4} J}\cosh\left(\tfrac{5}{4}\beta J\right)+{\rm e}^{3\beta J_1}+2{\rm e}^{\tfrac{\beta}{4}(J+8J_1)}\cosh\left(\tfrac{3}{4}\beta J\right)\right]
\big\}.
\label{a1}
\end{align}
\section{\label{App B} Partially transposed reduced density matrix $\hat{\rho}_{\mu S_1}^{T_{\mu}}$ }
Partially transposed reduced density matrix $\hat{\rho}_{\mu S_1}^{T_{\mu}}$  has a block diagonal structure
\begin{align}
\hat{\rho}_{\mu S_1}^{T_{\mu}}=\left(
\begin{array}{cccccc}
\omega_{33} ^-& 0 & 0 & 0 & 0 & 0\\
0 & \omega_{11}^- & \omega_{42}^- & 0 & 0 & 0\\
0 & \omega_{24}^- & \omega_{22}^+ & 0 & 0 & 0\\
0 & 0 & 0 & \omega_{22}^- & \omega_{24}^+ & 0 \\
0 & 0 & 0 & \omega_{42}^+ & \omega_{11}^+ & 0 \\
0 & 0 & 0 & 0 & 0 & \omega_{33}^+
\end{array}
\right),
\label{a2}
\end{align}%
where the non-zero elements are
\begin{align}
\omega_{11}^{\mp}&=\frac{1}{30{\cal Z}}\{
30\exp(-\beta\varepsilon_{\tfrac{5}{2},\pm\tfrac{5}{2}})
+12\exp(-\beta\varepsilon_{\tfrac{5}{2},\pm\tfrac{3}{2}})
+3\exp(-\beta\varepsilon_{\tfrac{5}{2},\pm\tfrac{1}{2}})
+3\exp(-\beta\varepsilon_{\tfrac{3}{2},\pm\tfrac{3}{2}}^{\rm II})+15\exp(-\beta\varepsilon_{\tfrac{3}{2},\pm\tfrac{3}{2}}^{\rm I})\nonumber\\
&+2\exp(-\beta\varepsilon_{\tfrac{3}{2},\pm\tfrac{1}{2}}^{\rm II})
+10\exp(-\beta\varepsilon_{\tfrac{3}{2},\pm\tfrac{1}{2}}^{\rm I})+5\exp(-\beta\varepsilon_{\tfrac{1}{2},\pm\tfrac{1}{2}}^{\rm II})+10\exp(-\beta\varepsilon_{\tfrac{1}{2},\pm\tfrac{1}{2}}^{\rm I})\}
\nonumber\\
&=\frac{{\rm e}^{-\beta J_1}}{30{\cal Z}}\big\{
30{\rm e}^{\pm\tfrac{5}{2}\beta h}{\rm e}^{-\beta J}
+3{\rm e}^{\pm\tfrac{3}{2}\beta h}[{\rm e}^{\tfrac{3}{2}\beta J}+5{\rm e}^{-\tfrac{\beta}{2}(J-4J_1)}+4{\rm e}^{-\beta J}]
\nonumber\\
&+{\rm e}^{\pm\tfrac{\beta}{2}h}[2{\rm e}^{\tfrac{3}{2}\beta J}+10{\rm e}^{3\beta J_1}+5{\rm e}^{\beta(J+2J_1)}+10{\rm e}^{-\tfrac{\beta}{2}(J-4J_1)}+3{\rm e}^{-\beta J}]
\big\},\nonumber\\
\omega_{22}^{\mp}&=\frac{1}{30{\cal Z}}
\{
12\exp(-\beta\varepsilon_{\tfrac{5}{2},\pm\tfrac{3}{2}})
+12\exp(-\beta\varepsilon_{\tfrac{5}{2},\pm\tfrac{1}{2}})
+6\exp(-\beta\varepsilon_{\tfrac{5}{2},\mp\tfrac{1}{2}})
+3\exp(-\beta\varepsilon_{\tfrac{3}{2},\pm\tfrac{3}{2}}^{\rm II})+15\exp(-\beta\varepsilon_{\tfrac{3}{2},\pm\tfrac{3}{2}}^{\rm I})\nonumber\\
&+8\exp(-\beta\varepsilon_{\tfrac{3}{2},\pm\tfrac{1}{2}}^{\rm II})
+9\exp(-\beta\varepsilon_{\tfrac{3}{2},\mp\tfrac{1}{2}}^{\rm II})
+5\exp(-\beta\varepsilon_{\tfrac{3}{2},\mp\tfrac{1}{2}}^{\rm I})+10\exp(-\beta\varepsilon_{\tfrac{1}{2},\mp\tfrac{1}{2}}^{\rm II})
+10\exp(-\beta\varepsilon_{\tfrac{1}{2},\pm\tfrac{1}{2}}^{\rm I})\}
\nonumber\\
&=\frac{{\rm e}^{-\beta J_1}}{30{\cal Z}}\big\{
3{\rm e}^{\pm\tfrac{3}{2}\beta h}[{\rm e}^{\tfrac{3}{2}\beta J}+5{\rm e}^{-\tfrac{\beta}{2}(J-4J_1)}+4{\rm e}^{-\beta J}]
+2{\rm e}^{\pm\tfrac{\beta}{2} h}[4{\rm e}^{\tfrac{3}{2}\beta J}+5{\rm e}^{3\beta J_1}+6{\rm e}^{-\beta J}]
\nonumber\\
&+{\rm e}^{\mp\tfrac{\beta}{2}h}[9{\rm e}^{\tfrac{3}{2}\beta J}+10{\rm e}^{\beta(J+2J_1)}+5{\rm e}^{-\tfrac{\beta}{2}(J-4J_1)}+6{\rm e}^{-\beta J}]
\big\},\nonumber\\
\omega_{33}^{\mp}&=\frac{1}{30{\cal Z}}\{
6\exp(-\beta\varepsilon_{\tfrac{5}{2},\mp\tfrac{3}{2}})
+3\exp(-\beta\varepsilon_{\tfrac{5}{2},\pm\tfrac{1}{2}})
+6\exp(-\beta\varepsilon_{\tfrac{5}{2},\mp\tfrac{1}{2}})
+24\exp(-\beta\varepsilon_{\tfrac{3}{2},\mp\tfrac{3}{2}}^{\rm II})
+2\exp(-\beta\varepsilon_{\tfrac{3}{2},\pm\tfrac{1}{2}}^{\rm II})\nonumber\\
&
+9\exp(-\beta\varepsilon_{\tfrac{3}{2},\mp\tfrac{1}{2}}^{\rm II})
+10\exp(-\beta\varepsilon_{\tfrac{3}{2},\pm\tfrac{1}{2}}^{\rm I})+5\exp(-\beta\varepsilon_{\tfrac{3}{2},\mp\tfrac{1}{2}}^{\rm I})+5\exp(-\beta\varepsilon_{\tfrac{1}{2},\pm\tfrac{1}{2}}^{\rm II})
+10\exp(-\beta\varepsilon_{\tfrac{1}{2},\mp\tfrac{1}{2}}^{\rm II})\nonumber\\&
+10\exp(-\beta\varepsilon_{\tfrac{1}{2},\pm\tfrac{1}{2}}^{\rm I})\}
=\frac{{\rm e}^{-\beta J_1}}{30{\cal Z}}\big\{
{\rm e}^{\pm\tfrac{\beta}{2} h}[2{\rm e}^{\tfrac{3}{2}\beta J}+10{\rm e}^{3\beta J_1}+5{\rm e}^{\beta(J+2J_1)}+10{\rm e}^{-\tfrac{\beta}{2}(J-4J_1)}+3{\rm e}^{-\beta J}]
\nonumber\\
&+{\rm e}^{\mp\tfrac{\beta}{2}h}[9{\rm e}^{\tfrac{3}{2}\beta J}+10{\rm e}^{\beta(J+2J_1)}+5{\rm e}^{-\tfrac{\beta}{2}(J-4J_1)}+6{\rm e}^{-\beta J}]
+6{\rm e}^{\mp\tfrac{3}{2}\beta h}[4{\rm e}^{\tfrac{3}{2}\beta J}+{\rm e}^{-\beta J}]
\big\},\nonumber\\
\omega_{24}^{\mp}&=\omega_{42}^{\mp}=\frac{\sqrt{2}}{30{\cal Z}}
\{
6\exp(-\beta\varepsilon_{\tfrac{5}{2},\pm\tfrac{3}{2}})
+6\exp(-\beta\varepsilon_{\tfrac{5}{2},\pm\tfrac{1}{2}})
+3\exp(-\beta\varepsilon_{\tfrac{5}{2},\mp\tfrac{1}{2}})
-6\exp(-\beta\varepsilon_{\tfrac{3}{2},\pm\tfrac{3}{2}}^{\rm II})
-6\exp(-\beta\varepsilon_{\tfrac{3}{2},\pm\tfrac{1}{2}}^{\rm II})\nonumber\\
&
-3\exp(-\beta\varepsilon_{\tfrac{3}{2},\mp\tfrac{1}{2}}^{\rm II})
+5\exp(-\beta\varepsilon_{\tfrac{3}{2},\mp\tfrac{1}{2}}^{\rm I})-5\exp(-\beta\varepsilon_{\tfrac{1}{2},\mp\tfrac{1}{2}}^{\rm II})\}
=\frac{\sqrt{2}{\rm e}^{-\beta J_1}}{30{\cal Z}}\big\{
-6{\rm e}^{\pm\tfrac{3}{2}\beta h}[{\rm e}^{\tfrac{3}{2}\beta J}-{\rm e}^{-\beta J}]
\nonumber\\
&
-6{\rm e}^{\pm\tfrac{\beta}{2}h}[{\rm e}^{\tfrac{3}{2}\beta J}-{\rm e}^{-\beta J}]-{\rm e}^{\mp\tfrac{\beta}{2}h}[3{\rm e}^{\tfrac{3}{2}\beta J}+5{\rm e}^{\beta(J+2J_1)}-5{\rm e}^{-\tfrac{\beta}{2}(J-4J_1)}-3{\rm e}^{-\beta J}]\big\}.
\label{a3}
\end{align}

Respective eigenvalues  can be read as follows:
\begin{align}
(\lambda_{1}^{\mp})_{\mu S_1}&=\omega_{33}^{\mp},\nonumber\\
(\lambda_{2}^{\mp})_{\mu S_1}&=\tfrac{1}{2}\left\{\omega_{11}^{\mp}+\omega_{22}^{\pm}-\sqrt{(\omega_{11}^{\mp}+\omega_{22}^{\pm})^2-4[\omega_{11}^{\mp}\omega_{22}^{\mp}-(\omega_{24}^{\mp})^2]} \right\},\nonumber\\
(\lambda_{3}^{\mp})_{\mu S_1}&=\tfrac{1}{2}\left\{\omega_{11}^{\mp}+\omega_{22}^{\pm}+\sqrt{(\omega_{11}^{\mp}+\omega_{22}^{\pm})^2-4[\omega_{11}^{\mp}\omega_{22}^{\mp}-(\omega_{24}^{\mp})^2]} \right\}.
\label{a4}
\end{align}
In general, only the eigenvalue $(\lambda_2^{\mp})_{\mu S_1}$ can be negative.

\section{\label{App C}Partially transposed reduced density matrix  $\hat{\rho}_{S_1S_2}^{T_{S_1}}$}
Partially transposed reduced density matrix $\hat{\rho}_{S_1S_2}^{T_{S_1}}$ has again a block diagonal structure
\begin{align}
\hat{\rho}_{S_1S_2}^{T_{S_1}}=\left(
\begin{array}{ccccccccc}
\mu_{33}& 0 & 0 & 0 & 0 & 0& 0 & 0 & 0\\
0 & \mu_{22}^- & \mu_{53} & 0 & 0 & 0& 0 & 0 & 0\\
0 & \mu_{35}& \mu_{22} ^+& 0 & 0 & 0& 0 & 0 & 0\\
0 & 0 & 0 & \mu_{11} ^-& \mu_{42}^- & \mu_{73}& 0 & 0 & 0 \\
0 & 0 & 0 & \mu_{24}^- & \mu_{55} & \mu_{24}^+& 0 & 0 & 0 \\
0 & 0 & 0 & \mu_{37} & \mu_{42}^+ & \mu_{11}^+& 0 & 0 & 0 \\
0 & 0 & 0& 0 & 0 & 0 & \mu_{22} ^-& \mu_{35}& 0\\
0 & 0 & 0& 0 & 0 & 0 & \mu_{53} & \mu_{22}^+ & 0\\
0 & 0 & 0 & 0 & 0& 0 & 0 & 0 & \mu_{33}
\end{array}
\right),
\label{b1}
\end{align}%
where the non-zero elements are
\begin{align}
\mu_{11}^{\mp}&=\frac{1}{5{\cal Z}}\{
5\exp(-\beta\varepsilon_{\tfrac{5}{2},\pm\tfrac{5}{2}})
+\exp(-\beta\varepsilon_{\tfrac{5}{2},\pm\tfrac{3}{2}})
+4\exp(-\beta\varepsilon_{\tfrac{3}{2},\pm\tfrac{3}{2}}^{\rm II})\}
=\frac{{\rm e}^{-\beta J_1}}{5{\cal Z}}\big\{
5{\rm e}^{\pm\tfrac{5}{2}\beta h}{\rm e}^{-\beta J}
+{\rm e}^{\pm\tfrac{3}{2}\beta h}[4{\rm e}^{\tfrac{3}{2}\beta J}+{\rm e}^{-\beta J}]\big\},
\nonumber\\
\mu_{22}^{\mp}&=\frac{1}{30{\cal Z}}\{
12\exp(-\beta\varepsilon_{\tfrac{5}{2},\pm\tfrac{3}{2}})
+6\exp(-\beta\varepsilon_{\tfrac{5}{2},\pm\tfrac{1}{2}})
+3\exp(-\beta\varepsilon_{\tfrac{3}{2},\pm\tfrac{3}{2}}^{\rm II})+15\exp(-\beta\varepsilon_{\tfrac{3}{2},\pm\tfrac{3}{2}}^{\rm I})+9\exp(-\beta\varepsilon_{\tfrac{3}{2},\pm\tfrac{1}{2}}^{\rm II})\nonumber\\
&
+5\exp(-\beta\varepsilon_{\tfrac{3}{2},\pm\tfrac{1}{2}}^{\rm I})+10\exp(-\beta\varepsilon_{\tfrac{1}{2},\pm\tfrac{1}{2}}^{\rm II})\}
\nonumber\\
&=\frac{{\rm e}^{-\beta J_1}}{30{\cal Z}}\big\{
3{\rm e}^{\pm\tfrac{3}{2}\beta h}[{\rm e}^{\tfrac{3}{2}\beta J}+5{\rm e}^{-\tfrac{\beta}{2}(J-4J_1)}+4{\rm e}^{-\beta J}]
+{\rm e}^{\pm\tfrac{\beta}{2}h}[9{\rm e}^{\tfrac{3}{2}\beta J}+10{\rm e}^{\beta(J+2J_1)}+5{\rm e}^{-\tfrac{\beta}{2}(J-4J_1)}+6{\rm e}^{-\beta J}]
\big\},\nonumber\\
\mu_{33}&=\frac{1}{30{\cal Z}}\{
3\exp(-\beta\varepsilon_{\tfrac{5}{2},+\tfrac{1}{2}})
+3\exp(-\beta\varepsilon_{\tfrac{5}{2},-\tfrac{1}{2}})
+2\exp(-\beta\varepsilon_{\tfrac{3}{2},+\tfrac{1}{2}}^{\rm II})+2\exp(-\beta\varepsilon_{\tfrac{3}{2},-\tfrac{1}{2}}^{\rm II})+10\exp(-\beta\varepsilon_{\tfrac{3}{2},+\tfrac{1}{2}}^{\rm I})\nonumber\\
&
+10\exp(-\beta\varepsilon_{\tfrac{3}{2},-\tfrac{1}{2}}^{\rm I})+5\exp(-\beta\varepsilon_{\tfrac{1}{2},+\tfrac{1}{2}}^{\rm II})+5\exp(-\beta\varepsilon_{\tfrac{1}{2},-\tfrac{1}{2}}^{\rm II})+10\exp(-\beta\varepsilon_{\tfrac{1}{2},+\tfrac{1}{2}}^{\rm I})+10\exp(-\beta\varepsilon_{\tfrac{1}{2},-\tfrac{1}{2}}^{\rm I})\}
\nonumber\\
&=\frac{{\rm e}^{-\beta J_1}}{30{\cal Z}}\big\{
2\cosh\left(\tfrac{\beta}{2}h\right)[2{\rm e}^{\tfrac{3}{2}\beta J}+10{\rm e}^{3\beta J_1}+5{\rm e}^{\beta(J+2J_1)}+10{\rm e}^{-\tfrac{\beta}{2}(J-4J_1)}+3{\rm e}^{-\beta J}]
\big\},\nonumber\\
\mu_{55}&=\frac{1}{15{\cal Z}}\{
6\exp(-\beta\varepsilon_{\tfrac{5}{2},+\tfrac{1}{2}})
+6\exp(-\beta\varepsilon_{\tfrac{5}{2},-\tfrac{1}{2}})
+4\exp(-\beta\varepsilon_{\tfrac{3}{2},+\tfrac{1}{2}}^{\rm II})+4\exp(-\beta\varepsilon_{\tfrac{3}{2},-\tfrac{1}{2}}^{\rm II})+5\exp(-\beta\varepsilon_{\tfrac{1}{2},+\tfrac{1}{2}}^{\rm I})\nonumber\\
&
+5\exp(-\beta\varepsilon_{\tfrac{1}{2},-\tfrac{1}{2}}^{\rm I})\}
=\frac{{\rm e}^{-\beta J_1}}{15{\cal Z}}\big\{
2\cosh\left(\tfrac{\beta}{2}h\right)[4{\rm e}^{\tfrac{3}{2}\beta J}+5{\rm e}^{3\beta J_1}+6{\rm e}^{-\beta J}]
\big\},\nonumber\\
\mu_{24}^{\mp}&=\mu_{42}^{\mp}=\frac{1}{30{\cal Z}}\{
12\exp(-\beta\varepsilon_{\tfrac{5}{2},\pm\tfrac{3}{2}})
+6\exp(-\beta\varepsilon_{\tfrac{5}{2},\pm\tfrac{1}{2}})
+3\exp(-\beta\varepsilon_{\tfrac{3}{2},\pm\tfrac{3}{2}}^{\rm II})-15\exp(-\beta\varepsilon_{\tfrac{3}{2},\pm\tfrac{3}{2}}^{\rm I})+9\exp(-\beta\varepsilon_{\tfrac{3}{2},\pm\tfrac{1}{2}}^{\rm II})\nonumber\\
&
-5\exp(-\beta\varepsilon_{\tfrac{3}{2},\pm\tfrac{1}{2}}^{\rm I})-10\exp(-\beta\varepsilon_{\tfrac{1}{2},\pm\tfrac{1}{2}}^{\rm II})\}\nonumber\\
&=\frac{{\rm e}^{-\beta J_1}}{30{\cal Z}}\big\{
3{\rm e}^{\pm\tfrac{3}{2}\beta h}[{\rm e}^{\tfrac{3}{2}\beta J}-5{\rm e}^{-\tfrac{\beta}{2}(J-4J_1)}+4{\rm e}^{-\beta J}]
+{\rm e}^{\pm\tfrac{\beta}{2}h}[9{\rm e}^{\tfrac{3}{2}\beta J}-10{\rm e}^{\beta(J+2J_1)}-5{\rm e}^{-\tfrac{\beta}{2}(J-4J_1)}+6{\rm e}^{-\beta J}]
\big\},\nonumber\\
\mu_{35}&=\mu_{53}=\frac{1}{15{\cal Z}}\{
3\exp(-\beta\varepsilon_{\tfrac{5}{2},+\tfrac{1}{2}})
+3\exp(-\beta\varepsilon_{\tfrac{5}{2},-\tfrac{1}{2}})
+2\exp(-\beta\varepsilon_{\tfrac{3}{2},+\tfrac{1}{2}}^{\rm II})+2\exp(-\beta\varepsilon_{\tfrac{3}{2},-\tfrac{1}{2}}^{\rm II})-5\exp(-\beta\varepsilon_{\tfrac{1}{2},\pm\tfrac{1}{2}}^{\rm I})\nonumber\\
&
-5\exp(-\beta\varepsilon_{\tfrac{1}{2},\pm\tfrac{1}{2}}^{\rm I})\}
=\frac{{\rm e}^{-\beta J_1}}{15{\cal Z}}\big\{
2\cosh\left(\tfrac{\beta}{2}h\right)[2{\rm e}^{\tfrac{3}{2}\beta J}-5{\rm e}^{3\beta J_1}+3{\rm e}^{-\beta J}]
\big\},\nonumber\\
\mu_{37}&=\mu_{73}=\frac{1}{30{\cal Z}}\{
3\exp(-\beta\varepsilon_{\tfrac{5}{2},+\tfrac{1}{2}})
+3\exp(-\beta\varepsilon_{\tfrac{5}{2},-\tfrac{1}{2}})
+2\exp(-\beta\varepsilon_{\tfrac{3}{2},+\tfrac{1}{2}}^{\rm II})+2\exp(-\beta\varepsilon_{\tfrac{3}{2},-\tfrac{1}{2}}^{\rm II})-10\exp(-\beta\varepsilon_{\tfrac{3}{2},+\tfrac{1}{2}}^{\rm I})\nonumber\\
&-10\exp(-\beta\varepsilon_{\tfrac{3}{2},-\tfrac{1}{2}}^{\rm I})
-5\exp(-\beta\varepsilon_{\tfrac{1}{2},+\tfrac{1}{2}}^{\rm II})\}-5\exp(-\beta\varepsilon_{\tfrac{1}{2},-\tfrac{1}{2}}^{\rm II})\}
+10\exp(-\beta\varepsilon_{\tfrac{1}{2},+\tfrac{1}{2}}^{\rm I})\}+10\exp(-\beta\varepsilon_{\tfrac{1}{2},-\tfrac{1}{2}}^{\rm I})\}
\nonumber\\
&=\frac{{\rm e}^{-\beta J_1}}{30{\cal Z}}\big\{
2\cosh\left(\tfrac{\beta}{2}h\right)[2{\rm e}^{\tfrac{3}{2}\beta J}+10{\rm e}^{3\beta J_1}-5{\rm e}^{\beta(J+2J_1)}-10{\rm e}^{-\tfrac{\beta}{2}(J-4J_1)}+3{\rm e}^{-\beta J}]
\big\}.
\label{b2}
\end{align}
Respective eigenvalues can be read as follows:
\begin{align}
(\lambda_{1})_{S_1S_2}&=\mu_{33},
\nonumber\\
(\lambda_{2})_{S_1S_2}&=\tfrac{1}{2}\left\{\mu_{22}^{-}+\mu_{22}^{+}-\sqrt{(\mu_{22}^{-}-\mu_{22}^{+})^2+4(\mu_{35})^2} \right\},
\nonumber\\
(\lambda_{3})_{S_1S_2}&=\tfrac{1}{2}\left\{\mu_{22}^{-}+\mu_{22}^{+}+\sqrt{(\mu_{22}^{-}-\mu_{22}^{+})^2+4(\mu_{35})^2} \right\},
\nonumber\\
(\lambda_{4,5,6})_{S_1S_2}&=\tfrac{a}{3}+2{\rm sign}(q)\sqrt{p}\cos(\tfrac{\phi}{3}+\tfrac{2\pi i}{3}),\;\;\;i=1,2,3
\label{a6}
\end{align}
where 
\begin{align}
p&=\left(\tfrac{a}{3}\right)^2-\tfrac{b}{3},\;\; q=\left(\tfrac{a}{3}\right)^3-\tfrac{a}{3}\tfrac{b}{2}-\tfrac{c}{2},\;\;
\phi=\arctan\left(\tfrac{\sqrt{p^3-q^2}}{q} \right),
\nonumber\\
a&=\mu_{11}^-+\mu_{11}^++\mu_{55},\;\;
\nonumber\\
b&=\mu_{11}^-(\mu_{11}^++\mu_{55})+\mu_{11}^+\mu_{55}-(\mu_{24}^-)^2-(\mu_{24}^+)^2-(\mu_{37})^2,
\nonumber\\
c&=-\mu_{11}^-\mu_{11}^+\mu_{55}+\mu_{11}^-(\mu_{24}^+)^2+\mu_{11}^+(\mu_{24}^-)^2+\mu_{55}(\mu_{37})^2-2\mu_{24}^-\mu_{24}^+\mu_{37}.
\label{a7}
\end{align}
%
It should be emphasized that each of the eigenvalues $(\lambda_1)_{S_1S_2}$, $(\lambda_2)_{S_1S_2}$, and $(\lambda_3)_{S_1S_2}$ is twofold degenerate.

\section{\label{App D}Density matrices $\hat{\rho}_{\mu S_1S_2}$}

The density matrix $\hat{\rho}_{\mu S_1S_2}$ defined in a standard basis $\{\vert \mu^z,S_1^z,S_2^z\rangle\}$ exhibits a block diagonal structure 
\begin{align}
\hat{\rho}_{\mu S_1S_2}=\left(
\begin{array}{cccccc}
A_1^- &     &     &       &     &\\
      &A_2^-&     &    0   &     &\\
      &     &A_3^-&       &     &\\
      &     &     & A_3^+ &     &\\
      &     &  0   &       &A_2^+&  \\
      &     &     &       &     & A_1^+\\
\end{array}
\right),
\label{c1}
\end{align}
consisting of the following six blocks
\begin{align}
A_1^{\mp}&=\left(
\begin{array}{c}
\rho_{1,1}^{\mp}
\end{array}
\right),\;
A_2^{\mp}=\left(
\begin{array}{ccc}
\rho_{2,2}^{\mp} & \rho_{2,4}^{\mp}  & \rho_{2,10}^{\mp} \\
\rho_{4,2}^{\mp} & \rho_{2,2}^{\mp}  & \rho_{2,10}^{\mp} \\
\rho_{10,2}^{\mp} & \rho_{10,2}^{\mp}  & \rho_{9,9}^{\mp} 
\end{array}
\right),
\;
A_3^{\mp}=\left(
\begin{array}{ccccc}
\rho_{3,3}^{\mp} & \rho_{3,5}^{\mp}  & \rho_{3,7}^{\mp} & \rho_{3,11}^{\mp} & \rho_{3,13}^{\mp} \\
\rho_{5,3}^{\mp} & \rho_{5,5}^{\mp}  & \rho_{3,5}^{\mp} & \rho_{5,11}^{\mp} & \rho_{5,11}^{\mp} \\
\rho_{7,3}^{\mp} & \rho_{5,3}^{\mp}  & \rho_{3,3}^{\mp} & \rho_{3,13}^{\mp} & \rho_{3,11}^{\mp} \\
\rho_{11,3}^{\mp} & \rho_{11,5}^{\mp}  & \rho_{13,3}^{\mp} & \rho_{6,6}^{\mp} & \rho_{6,8}^{\mp} \\
\rho_{13,3}^{\mp} & \rho_{11,5}^{\mp}  & \rho_{11,3}^{\mp} & \rho_{8,6}^{\mp} & \rho_{6,6}^{\mp} 
\end{array}
\right).
\label{a8}
\end{align}%
The non-zero density matrix elements are expressed analytically as follows:
\begin{align}
\rho_{1,1}^{\mp}&=\frac{1}{\cal Z}\exp(-\beta\varepsilon_{\tfrac{5}{2},\pm\tfrac{5}{2}})=\frac{{\rm e}^{-\beta J_1}}{\cal Z}{\rm e}^{\pm\frac{5}{2}\beta h}{\rm e}^{-\beta J},
\nonumber\\
\rho_{2,2}^{\mp}&=\frac{1}{10\cal Z}\{
4\exp(\beta\varepsilon_{\tfrac{5}{2},\pm\tfrac{3}{2}})
+\exp(-\beta\varepsilon_{\tfrac{3}{2},\pm\tfrac{3}{2}}^{\rm II})+5\exp(-\beta\varepsilon_{\tfrac{3}{2},\pm\tfrac{3}{2}}^{\rm I})\}
\nonumber\\
&=\frac{{\rm e}^{-\beta J_1}}{10\cal Z}{\rm e}^{\pm\frac{3}{2}\beta h}\big\{
{\rm e}^{\tfrac{3}{2}\beta J}+5{\rm e}^{-\tfrac{\beta}{2}(J-4J_1)}+4{\rm e}^{-\beta J}
\big\},\nonumber\\
\rho_{3,3}^{\mp}&=\frac{1}{30\cal Z}\{
3\exp(-\beta\varepsilon_{\tfrac{5}{2},\pm\tfrac{1}{2}})
+2\exp(-\beta\varepsilon_{\tfrac{3}{2},\pm\tfrac{1}{2}}^{\rm II})+10\exp(-\beta\varepsilon_{\tfrac{3}{2},\pm\tfrac{1}{2}}^{\rm I})
+5\exp(-\beta\varepsilon_{\tfrac{1}{2},\pm\tfrac{1}{2}}^{\rm II})+10\exp(-\beta\varepsilon_{\tfrac{1}{2},\pm\tfrac{1}{2}}^{\rm I})\}
\nonumber\\
&=\frac{{\rm e}^{-\beta J_1}}{30{\cal Z}}{\rm e}^{\pm\tfrac{\beta}{2} h}\big\{
2{\rm e}^{\tfrac{3}{2}\beta J}+10{\rm e}^{3\beta J_1}+5{\rm e}^{\beta(J+2J_1)}+10{\rm e}^{-\tfrac{\beta}{2}(J-4J_1)}+3{\rm e}^{-\beta J}
\big\},\nonumber\\
%
%
\rho_{5,5}^{\mp}&=\frac{1}{15\cal Z}\{
6\exp(-\beta\varepsilon_{\tfrac{5}{2},\pm\tfrac{1}{2}})
+4\exp(-\beta\varepsilon_{\tfrac{3}{2},\pm\tfrac{1}{2}}^{\rm II})+5\exp(-\beta\varepsilon_{\tfrac{1}{2},\pm\tfrac{1}{2}}^{\rm I})\}
=\frac{{\rm e}^{-\beta J_1}}{15{\cal Z}}{\rm e}^{\pm\tfrac{\beta}{2} h}\big\{
4{\rm e}^{\tfrac{3}{2}\beta J}+5{\rm e}^{3\beta J_1}+6{\rm e}^{-\beta J}
\big\},\nonumber\\
\rho_{6,6}^{\mp}&=\frac{1}{30\cal Z}\{
6\exp(-\beta\varepsilon_{\tfrac{5}{2},\pm\tfrac{1}{2}})
+9\exp(-\beta\varepsilon_{\tfrac{3}{2},\pm\tfrac{1}{2}}^{\rm II})+5\exp(-\beta\varepsilon_{\tfrac{3}{2},\pm\tfrac{1}{2}}^{\rm I})
+10\exp(-\beta\varepsilon_{\tfrac{1}{2},\pm\tfrac{1}{2}}^{\rm II})\}
\nonumber\\
&=\frac{{\rm e}^{-\beta J_1}}{30{\cal Z}}{\rm e}^{\pm\tfrac{\beta}{2} h}\big\{
9{\rm e}^{\tfrac{3}{2}\beta J}+10{\rm e}^{\beta(J+2J_1)}+5{\rm e}^{-\tfrac{\beta}{2}(J-4J_1)}+6{\rm e}^{-\beta J}
\big\},\nonumber\\
%
%
%
\rho_{9,9}^{\mp}&=\frac{1}{5\cal Z}\{
\exp(-\beta\varepsilon_{\tfrac{5}{2},\pm\tfrac{3}{2}})
+4\exp(-\beta\varepsilon_{\tfrac{3}{2},\pm\tfrac{3}{2}}^{\rm II})\}
=\frac{{\rm e}^{-\beta J_1}}{5{\cal Z}}{\rm e}^{\pm\tfrac{3\beta}{2} h}\big\{
4{\rm e}^{\tfrac{3}{2}\beta J}+{\rm e}^{-\beta J}
\big\},\nonumber\\
\rho_{2,4}^{\mp}&=\rho_{4,2}^{\mp}=\frac{1}{10\cal Z}\{
4\exp(-\beta\varepsilon_{\tfrac{5}{2},\pm\tfrac{3}{2}})
+\exp(-\beta\varepsilon_{\tfrac{3}{2},\pm\tfrac{3}{2}}^{\rm II})-5\exp(-\beta\varepsilon_{\tfrac{3}{2},\pm\tfrac{3}{2}}^{\rm I})\}
\nonumber\\
&=\frac{{\rm e}^{-\beta J_1}}{10{\cal Z}}{\rm e}^{\pm\tfrac{3\beta}{2} h}\big\{
{\rm e}^{\tfrac{3}{2}\beta J}-5{\rm e}^{-\tfrac{\beta}{2}(J-4J_1)}+4{\rm e}^{-\beta J}
\big\},\nonumber\\
\rho_{2,10}^{\mp}&=\rho_{10,2}^{\mp}=\frac{\sqrt{2}}{5\cal Z}\{
\exp(-\beta\varepsilon_{\tfrac{5}{2},\pm\tfrac{3}{2}})
-\exp(-\beta\varepsilon_{\tfrac{3}{2},\pm\tfrac{3}{2}}^{\rm II})\}
=-\frac{\sqrt{2}}{5{\cal Z}}{\rm e}^{-\beta J_1}{\rm e}^{\pm\tfrac{3\beta}{2} h}\big\{
{\rm e}^{\tfrac{3}{2}\beta J}-{\rm e}^{-\beta J}
\big\},\nonumber\\
\rho_{3,5}^{\mp}&=\rho_{5,3}^{\mp}=\frac{1}{15\cal Z}\{
3\exp(-\beta\varepsilon_{\tfrac{5}{2},\pm\tfrac{1}{2}})
+2\exp(-\beta\varepsilon_{\tfrac{3}{2},\pm\tfrac{1}{2}}^{\rm II})-5\exp(-\beta\varepsilon_{\tfrac{1}{2},\pm\tfrac{1}{2}}^{\rm I})\}
\nonumber\\
&=\frac{{\rm e}^{-\beta J_1}}{15{\cal Z}}{\rm e}^{\pm\tfrac{\beta}{2} h}\big\{
2{\rm e}^{\tfrac{3}{2}\beta J}-5{\rm e}^{3\beta J_1}+3{\rm e}^{-\beta J}
\big\},\nonumber\\
\rho_{3,7}^{\mp}&=\rho_{7,3}^{\mp}=\frac{1}{30\cal Z}\{
3\exp(-\beta\varepsilon_{\tfrac{5}{2},\pm\tfrac{1}{2}})
+2\exp(-\beta\varepsilon_{\tfrac{3}{2},\pm\tfrac{1}{2}}^{\rm II})-10\exp(-\beta\varepsilon_{\tfrac{3}{2},\pm\tfrac{1}{2}}^{\rm I})
-5\exp(-\beta\varepsilon_{\tfrac{1}{2},\pm\tfrac{1}{2}}^{\rm II})
\}
\nonumber\\
&+10\exp(-\beta\varepsilon_{\tfrac{1}{2},\pm\tfrac{1}{2}}^{\rm I})=\frac{{\rm e}^{-\beta J_1}}{30{\cal Z}}{\rm e}^{\pm\tfrac{\beta}{2} h}\big\{
2{\rm e}^{\tfrac{3}{2}\beta J}+10{\rm e}^{3\beta J_1}-5{\rm e}^{\beta(J+2J_1)}-10{\rm e}^{-\tfrac{\beta}{2}(J-4J_1)}+3{\rm e}^{-\beta J}
\big\}
\nonumber\\
%
\rho_{3,11}^{\mp}&=\rho_{11,3}^{\mp}=-\frac{\sqrt{2}}{30\cal Z}\{
-3\exp(-\beta\varepsilon_{\tfrac{5}{2},\pm\tfrac{1}{2}})
+3\exp(-\beta\varepsilon_{\tfrac{3}{2},\pm\tfrac{1}{2}}^{\rm II})
-5\exp(-\beta\varepsilon_{\tfrac{3}{2},\pm\tfrac{1}{2}}^{\rm I})+5\exp(-\beta\varepsilon_{\tfrac{1}{2},\pm\tfrac{1}{2}}^{\rm II})\}
\nonumber\\
&=-\frac{\sqrt{2}}{30{\cal Z}}{\rm e}^{-\beta J_1}{\rm e}^{\pm\tfrac{\beta}{2} h}\big\{
3{\rm e}^{\tfrac{3}{2}\beta J}+5{\rm e}^{\beta(J+2J_1)}-5{\rm e}^{-\tfrac{\beta}{2}(J-4J_1)}-3{\rm e}^{-\beta J}
\big\},\nonumber\\
\rho_{3,13}^{\mp}&=\rho_{13,3}^{\mp}=-\frac{\sqrt{2}}{30\cal Z}\{
-3\exp(-\beta\varepsilon_{\tfrac{5}{2},\pm\tfrac{1}{2}})
+3\exp(-\beta\varepsilon_{\tfrac{3}{2},\pm\tfrac{1}{2}}^{\rm II})
+5\exp(-\beta\varepsilon_{\tfrac{3}{2},\pm\tfrac{1}{2}}^{\rm I})-5\exp(-\beta\varepsilon_{\tfrac{1}{2},\pm\tfrac{1}{2}}^{\rm II})\}
\nonumber\\
&=-\frac{\sqrt{2}}{30{\cal Z}}{\rm e}^{-\beta J_1}{\rm e}^{\pm\tfrac{\beta}{2} h}\big\{
3{\rm e}^{\tfrac{3}{2}\beta J}-5{\rm e}^{\beta(J+2J_1)}+5{\rm e}^{-\tfrac{\beta}{2}(J-4J_1)}-3{\rm e}^{-\beta J}
\big\},\nonumber\\
%
%
\rho_{5,11}^{\mp}&=\rho_{11,5}^{\mp}=\frac{\sqrt{2}}{5\cal Z}\{
\exp(-\beta\varepsilon_{\tfrac{5}{2},\pm\tfrac{1}{2}})
-\exp(-\beta\varepsilon_{\tfrac{3}{2},\pm\tfrac{1}{2}}^{\rm II})
\}=-\frac{\sqrt{2}}{5{\cal Z}}{\rm e}^{-\beta J_1}{\rm e}^{\pm\tfrac{\beta}{2} h}\big\{
{\rm e}^{\tfrac{3}{2}\beta J}-{\rm e}^{-\beta J}
\big\},\nonumber\\
%
%
\rho_{6,8}^{\mp}&=\rho_{8,6}^{\mp}=\frac{1}{30\cal Z}\{
6\exp(-\beta\varepsilon_{\tfrac{5}{2},\pm\tfrac{1}{2}})
+9\exp(-\beta\varepsilon_{\tfrac{3}{2},\pm\tfrac{1}{2}}^{\rm II})
-5\exp(-\beta\varepsilon_{\tfrac{3}{2},\pm\tfrac{1}{2}}^{\rm I})-10\exp(-\beta\varepsilon_{\tfrac{1}{2},\pm\tfrac{1}{2}}^{\rm II})\}
\nonumber\\
&=\frac{{\rm e}^{-\beta J_1}}{30{\cal Z}}{\rm e}^{\pm\tfrac{\beta}{2} h}\big\{
9{\rm e}^{\tfrac{3}{2}\beta J}-10{\rm e}^{\beta(J+2J_1)}-5{\rm e}^{-\tfrac{\beta}{2}(J-4J_1)}+6{\rm e}^{-\beta J}
\big\}.
%
%
%
\label{a9}
\end{align}

\section{\label{App E}Partially transposed density matrices $\hat{\rho}_{\mu S_1S_2}^{T_{\mu}}$ }

The partially transposed density matrix $\hat{\rho}^{T_\mu}_{\mu S_1S_2}$ also possesses a block diagonal structure, comprising two $1\times 1$ matrices $\mathbf{Q}_1^{\mu}(\mp)$, two $3\times 3$ matrices $\mathbf{Q}_2^{\mu}(\mp)$, and two $5\times 5$ matrices $\mathbf{Q}_3^{\mu}(\mp)$ with the following forms:
\begin{align}
\allowdisplaybreaks
\mathbf{Q}_1^{\mu}(\mp)\!&=\!\left(
\begin{array}{ccc}
 \rho_{9,9}^{\mp} 
\end{array}\right),\;\;
\mathbf{Q}_2^{\mu}(\mp)\!=\!\left(
\begin{array}{ccc}
 \rho_{1,1}^{\mp} & \rho_{2,10}^{\mp} & \rho_{2,10}^{\mp}\\
 \rho_{2,10}^{\mp} & \rho_{6,6}^{\mp} & \rho_{6,8}^{\mp}\\
 \rho_{2,10}^{\mp} & \rho_{6,8}^{\mp} & \rho_{6,6}^{\mp}\\
\end{array}\right),\;\;
\mathbf{Q}_3^{\mu}(\mp)\!=\!\left(
\begin{array}{ccccc}
 \rho_{2,2}^{\mp} & \rho_{2,4}^{\mp} & \rho_{3,11}^{\mp}& \rho_{5,11}^{\mp}& \rho_{3,13}^{\mp}\\
 \rho_{2,4}^{\mp} & \rho_{2,2}^{\mp} & \rho_{3,13}^{\mp}& \rho_{5,11}^{\mp}& \rho_{3,11}^{\mp}\\
 \rho_{3,11}^{\mp} & \rho_{3,13}^{\mp} & \rho_{3,3}^{\pm}& \rho_{3,5}^{\pm}& \rho_{3,7}^{\pm}\\
  \rho_{5,11}^{\mp} & \rho_{5,11}^{\mp} & \rho_{3,5}^{\pm}& \rho_{5,5}^{\pm}& \rho_{3,5}^{\pm}\\
  \rho_{3,13}^{\mp} & \rho_{3,11}^{\mp} & \rho_{3,7}^{\pm}& \rho_{3,5}^{\pm}& \rho_{3,3}^{\pm}
\end{array}\right).
\label{a12}
\end{align}
Respective eigenvalues are as follows:
\begin{align}
(\lambda_{1}^{\mp})_{\mu}&=\rho_{99}^{\mp},\nonumber\\
(\lambda_{2}^{\mp})_{\mu}&=\rho_{6,6}^{\mp}-\rho_{6,8}^{\mp},\nonumber\\
(\lambda_{3}^{\mp})_{\mu}&=\tfrac{1}{2}\left\{\rho_{1,1}^{\mp}+\rho_{66}^{\mp}+\rho_{6,8}^{\mp}-\sqrt{ (\rho_{1,1}^{\mp}-(\rho_{6,6}^{\mp}+\rho_{6,8}^{\mp}))^2+8(\rho_{2,10}^{\mp})^2 } 
\right\},\nonumber\\
(\lambda_{4}^{\mp})_{\mu}&=\tfrac{1}{2}\left\{\rho_{1,1}^{\mp}+\rho_{66}^{\mp}+\rho_{6,8}^{\mp}+\sqrt{ (\rho_{1,1}^{\mp}-(\rho_{6,6}^{\mp}+\rho_{6,8}^{\mp}))^2+8(\rho_{2,10}^{\mp})^2 } 
\right\},\nonumber\\
(\lambda_{5}^{\mp})_{\mu}&=\tfrac{1}{2}\left\{(\rho_{2,2}^{\mp}-\rho_{2,4}^{\mp})+(\rho_{3,3}^{\pm}-\rho_{3,7}^{\pm})-\sqrt{ ((\rho_{2,2}^{\mp}-\rho_{2,4}^{\mp})-(\rho_{3,3}^{\pm}-\rho_{3,7}^{\pm}))^2+4(\rho_{3,11}^{\mp}-\rho_{3,13}^{\mp})^2 } 
\right\},\nonumber\\
(\lambda_{6}^{\mp})_{\mu}&=\tfrac{1}{2}\left\{(\rho_{2,2}^{\mp}-\rho_{2,4}^{\mp})+(\rho_{3,3}^{\pm}-\rho_{3,7}^{\pm})+\sqrt{ ((\rho_{2,2}^{\mp}-\rho_{2,4}^{\mp})-(\rho_{3,3}^{\pm}-\rho_{3,7}^{\pm}))^2+4(\rho_{3,11}^{\mp}-\rho_{3,13}^{\mp})^2 } 
\right\},\nonumber\\
(\lambda_{7}^{\mp})_{\mu}&=\rho_{3,3}^{\pm}+\rho_{3,7}^{\pm}-\rho_{3,5}^{\pm},\nonumber\\
(\lambda_{8}^{\mp})_{\mu}&=\tfrac{1}{2}\left\{(\rho_{2,2}^{\mp}+\rho_{2,4}^{\mp})+(\rho_{5,5}^{\pm}+\rho_{3,5}^{\pm})-\sqrt{ ((\rho_{2,2}^{\mp}+\rho_{2,4}^{\mp})-(\rho_{5,5}^{\pm}+\rho_{3,5}^{\pm}))^2+12(\rho_{5,11}^{\mp})^2 } 
\right\},\nonumber\\
(\lambda_{9}^{\mp})_{\mu}&=\tfrac{1}{2}\left\{(\rho_{2,2}^{\mp}+\rho_{2,4}^{\mp})+(\rho_{5,5}^{\pm}+\rho_{3,5}^{\pm})+\sqrt{ ((\rho_{2,2}^{\mp}+\rho_{2,4}^{\mp})-(\rho_{5,5}^{\pm}+\rho_{3,5}^{\pm}))^2+12(\rho_{5,11}^{\mp})^2 } 
\right\}.
\label{a13}
\end{align}
Only six eigenvalues, $(\lambda_3^{\mp})_{\mu}$,  $(\lambda_5^{\mp})_{\mu}$, and  $(\lambda_8^{\mp})_{\mu}$ can generally be negative.

\section{\label{App F}Partially transposed density matrices  $\hat{\rho}_{\mu S_1S_2}^{T_{S_1}}$}

The partially transposed density matrix $\hat{\rho}^{T_{S_1}}_{\mu S_1S_2}$ also exhibits a block diagonal structure, composed of two  $1\times 1$ matrices $\mathbf{Q}_1^{S_1}(\mp)$, two  $3\times 3$ matrices $\mathbf{Q}_2^{S_1}(\mp)$ and two $5\times 5$ matrices $\mathbf{Q}_3^{S_1}(\mp)$  with the following forms:
\begin{align}
\allowdisplaybreaks
\mathbf{Q}_1^{S_1}(\mp)\!&=\!\left(
\begin{array}{ccc}
 \rho_{3,3}^{\mp} 
\end{array}\right),\;\;
\mathbf{Q}_2^{S_1}(\mp)\!=\!\left(
\begin{array}{ccc}
 \rho_{3,3}^{\mp} & \rho_{3,11}^{\mp} & \rho_{3,11}^{\pm}\\
 \rho_{3,11}^{\mp} & \rho_{6,6}^{\mp} & \rho_{3,5}^{\pm}\\
 \rho_{3,11}^{\pm} & \rho_{3,5}^{\pm} & \rho_{2,2}^{\pm}\\
\end{array}\right),\;\;
\mathbf{Q}_3^{S_1}(\mp)\!=\!\left(
\begin{array}{ccccc}
 \rho_{1,1}^{\mp} & \rho_{2,4}^{\mp} & \rho_{2,10}^{\mp}& \rho_{3,7}^{\mp}& \rho_{3,13}^{\mp}\\
 \rho_{2,4}^{\mp} & \rho_{5,5}^{\mp} & \rho_{5,11}^{\mp}& \rho_{6,8}^{\pm}& \rho_{5,11}^{\pm}\\
 \rho_{2,10}^{\mp} & \rho_{5,11}^{\mp} & \rho_{6,6}^{\mp}& \rho_{3,13}^{\pm}& \rho_{3,5}^{\pm}\\
  \rho_{3,7}^{\mp} & \rho_{6,8}^{\pm} & \rho_{3,13}^{\pm}& \rho_{9,9}^{\pm}& \rho_{2,10}^{\pm}\\
  \rho_{3,13}^{\mp} & \rho_{5,11}^{\pm} & \rho_{3,5}^{\pm}& \rho_{2,10}^{\pm}& \rho_{2,2}^{\pm}
\end{array}\right).
\label{a14}
\end{align}
For the first two blocks $\mathbf{Q}_1^{S_1}(\mp)$ and $\mathbf{Q}_2^{S_1}(\mp)$ we identify the following analytical expression of respective eigenvalues
\begin{align}
(\lambda_{1}^{\mp})_{S_1}&=\rho_{33}^{\mp},\nonumber\\
(\lambda_{2,3,4}^{\mp})_{S_1}&=\tfrac{a^{\mp}}{3}+2{\rm sign}(q^{\mp})\sqrt{p^{\mp}}\cos(\tfrac{\phi^{\mp}}{3}+\tfrac{2\pi i}{3}),\;\;\; i=1,2,3
\label{a15}
\end{align}
where 
\begin{align}
p^{\mp}&=\left(\tfrac{a^{\mp}}{3}\right)^2-\tfrac{b^{\mp}}{3},\;\; q^{\mp}=\left(\tfrac{a^{\mp}}{3}\right)^3-\tfrac{a^{\mp}}{3}\tfrac{b^{\mp}}{2}-\tfrac{c^{\mp}}{2},\;\;
\phi^{\mp}=\arctan\left(\tfrac{\sqrt{{p^{\mp}}^3-{q^{\mp}}^2}}{q^{\mp}} \right),
\nonumber\\
a^{\mp}&=\rho_{3,3}^{\mp}+\rho_{6,6}^{\mp}+\rho_{2,2}^{\pm},\;\;
\nonumber\\
b^{\mp}&=\rho_{3,3}^{\mp}(\rho_{6,6}^{\mp}+\rho_{2,2}^{\pm})+\rho_{6,6}^{\mp}\rho_{2,2}^{\pm}-(\rho_{3,11}^{\mp})^2-(\rho_{3,11}^{\pm})^2-(\rho_{3,5}^{\pm})^2,
\nonumber\\
c^{\mp}&=-\rho_{3,3}^{\mp}\rho_{6,6}^{\mp}\rho_{2,2}^{\pm}+\rho_{3,3}^{\mp}(\rho_{3,5}^{\pm})^2+\rho_{6,6}^{\mp}(\rho_{3,11}^{\pm})^2+\rho_{2,2}^{\pm}(\rho_{3,11}^{\mp})^2-2\rho_{3,11}^{\mp}\rho_{3,11}^{\pm}\rho_{3,5}^{\pm}.
\label{a16}
\end{align}
Unfortunately, the eigenvalues of the last block $\mathbf{Q}_3^{S_1}(\mp)$ can be determined only numerically.

\printcredits



\end{document}